\documentclass[nofootinbib,preprintnumbers,superscriptaddress]{revtex4-1}

\usepackage{amsmath}
\usepackage{mathrsfs} 
\usepackage{amssymb}
\usepackage{graphicx}
\usepackage{hyperref}
\usepackage{xspace}
\usepackage{bm}
\usepackage{slashed}
\usepackage{dsfont}

\usepackage{float} 
\usepackage{caption}
\usepackage{subcaption}

\usepackage{xcolor}
\definecolor{light-gray}{gray}{0.8}

\newcommand{\gS}{\mathrm{S}}
\newcommand{\gP}{\mathrm{P}}
\newcommand{\gV}{\mathrm{V}}
\newcommand{\gA}{\mathrm{A}}

\newcommand{\sym}{\mathrm{sym}}
\newcommand{\gam}[1]{\mathrm{#1}}
\newcommand{\chlim}{\lim_{m_R \to 0}}
\newcommand{\mlim}{\lim_{M_R \to \overline{m}}}
\newcommand{\dslash}{\not{\hbox{\kern-2pt $\partial$}}}
\newcommand{\pslash}{\not{\hbox{\kern-2pt $p$}}}
\newcommand{\kslash}{\not{\hbox{\kern-2pt $k$}}}
\newcommand{\qslash}{\not{\hbox{\kern-2pt $q$}}}
\newcommand{\mb}{\bar{m}}
 
\newcommand{\Tr}{\mbox{Tr}}
 
\newcommand{\sxi}{\sum_{i=1}^3 x_i}
\newcommand{\mmu}{\frac{m^2}{\mu^2}}
\newcommand{\mmmu}{\frac{m^4}{\mu^4}}
\newcommand{\plog}[1]{\log\left( #1\right)}

\newcommand\li{\text{ Li}}
\def\ie{{\it i.e.}\ }
\def\eg{{\it e.g.}\ }

\newcommand{\EDaff}{Higgs Centre for Theoretical Physics, School of Physics \& Astronomy,
  University of Edinburgh, Edinburgh EH9 3FD, United Kingdom.}

\begin{document}

\title{A massive momentum-subtraction scheme}

\preprint{xxx}

\author{Peter Boyle}
\affiliation{\EDaff}

\author{Luigi Del Debbio}
\affiliation{\EDaff}

\author{Ava Khamseh}
\affiliation{\EDaff}

\begin{abstract} 
A new renormalization scheme is defined for fermion bilinears in QCD
at non vanishing quark masses. This new scheme, denoted RI/mSMOM,
preserves the benefits of the nonexceptional momenta introduced in the
RI/SMOM scheme, and allows a definition of renormalized composite
fields away from the chiral limit. Some properties of the scheme are
investigated by performing explicit one-loop computation in
dimensional regularization. 
\end{abstract}

\pacs{}

\maketitle

\section{Introduction}
\label{sec:introduction}

Nonperturbative renormalization MOM schemes have been introduced in
Refs.~\cite{Martinelli:1994ty,Sturm:2009kb} by imposing a set of
renormalization conditions, which specify the renormalization of the
fermion wave function, of the fermion mass, and of composite operators
like fermion bilinears. The renormalization conditions are imposed in
the chiral limit of QCD, and therefore, by construction, these schemes
are mass-independent, meaning that all the renormalization constants
are independent of the value of the fermion mass. This is useful for
instance when considering ratios of quantities such as masses; in
a mass-independent scheme $m_i/m_j$ for two different fermions $i$ and
$j$, does not renormalize since the renormalization constants cancel
between the numerator and the denominator. The renormalization
conditions are chosen so that renormalized correlators involving the
vector and axial currents satisfy the Ward identities (WIs) dictated
by the symmetries of the theory. Using massless schemes for massive
quarks involves violations of the Ward identities by terms that scale
like powers of $m/\mu$, where $\mu$ is the typical energy scale of the
correlators that are computed. 

Recent lattice studies have begun investigating the nonperturbative
dynamics of heavy quarks like charm and bottom, including these heavy
flavors as relativistic dynamical degrees of freedom in the path
integral. In current simulations the mass of the heavy quarks is often
of the same order of magnitude as the UV cutoff, defined as the
inverse lattice spacing $a^{-1}$. As a consequence, it is not possible
to reach a regime where there is a clear separation between the
fermion mass, the renormalization scale, and the cutoff, \ie a regime
where $m \ll \mu \ll a^{-1}$. When studying heavy quarks, it may be
interesting to introduce a massive scheme, i.e. a scheme where the
renormalization conditions are imposed at some finite value of the
renormalized mass. It is indeed possible to choose the renormalization
conditions in such a way that the desirable properties of the massless
schemes are preserved, in particular the Ward identities would hold
exactly at finite values of the quark mass, and independently of the
ratio $m/\mu$.

In this paper, we define a massive scheme for axial and vector currents as well as scalar and pseudoscalar densities, which we call mSMOM. The
renormalization constants defined in mSMOM satisfy properties that are
similar to the ones found in SMOM~\cite{Sturm:2009kb}. SMOM was
introduced in order to reduce chiral symmetry breaking and other
unwanted infrared effects, by defining the renormalization conditions
for the vertex functions at a symmetric subtraction point which
involves non-exceptional momenta. The key property of the SMOM scheme
is that the renormalization conditions are defined so that the
renormalized WIs are satisfied. This is in contrast with MOM where the
WI for the axial current are recovered only for large values of
$\mu^2$ \cite{Sturm:2009kb}. Starting from SMOM, we modify some of the
renormalization conditions in order to recover the massive
renormalized WIs. The renormalization conditions for massive quarks
require the introduction of an extra scale $\overline{m}$, which is
the value of the renormalized mass at which the conditions are spelled
out. As we take the limit $\overline{m}\to 0$, our scheme reduces to
SMOM, so that we are able to interpolate between massive and massless
schemes.

We discuss a number of properties using non-perturbative arguments
after which we perform an explicit check at one-loop in perturbation
theory using dimensional regularization. While the results of this
calculation is exactly as expected, it is pleasing to see explicitly a
number of nontrivial cancellations. We then focus on the case of the
lattice currents, and discuss their renormalization in mSMOM. The
massive schemes can be implemented numerically, in order to obtain
nonperturbative determinations of the corresponding renormalization
constants. The massive renormalization constants will change some lattice artefacts $O(a^2m^2)$, and could potentially
lead to smoother extrapolations to the continuum limit of
phenomenologically relevant observables. A first qualitative
understanding of the can be obtained by a perturbative study along the
lines of Ref.~\cite{Athenodorou:2011zp}, but ultimately dedicated
numerical studies are necessary in order to settle this issue.

\section{Massive renormalization conditions}
\label{sec:renorm-cond}

A regularization independent momentum subtraction scheme for bilinears
with a nonexceptional, symmetric point has been introduced in
Ref.~\cite{Sturm:2009kb}, under the name of RI/SMOM. RI/SMOM is a
mass-independent renormalization scheme, in that all the
renormalization conditions are specified in the chiral limit, and
therefore the renormalization constants cannot depend on the quark
masses by definition. Before investigating the possibility of
defining a similar scheme at finite quark mass, let us briefly recall
the renormalization conditions that define RI/SMOM, and discuss the
main properties of the renormalized bilinears in that
scheme. 

Fig.~\ref{fig:kin} summarises the kinematics used in this
paper: the correlators of fermion bilinears with two
external off-shell fermions are
\begin{equation}
  \label{eq:Gcorr}
  G^a_\Gamma(p_3,p_2) = \langle O^a_\Gamma(q) \bar\psi(p_3) \psi(p_2)
  \rangle\, ,
\end{equation}
where $O^a_\Gamma=\bar\psi \Gamma\tau^a\psi$ is a flavor non-singlet
fermion bilinear, and $\Gamma$ spans all the elements of the basis of
the Clifford algebra, which we denote as
$\Gamma=\gS,\gP,\gV,\gA,\gam{T}$. Note that $\tau^a$ denotes a generic
generator of rotations in flavor space. The conventions for the Dirac
gamma matrices are spelled out in detail in
App.~\ref{sec:conventions}. The four dimensional vectors $p_2$ and
$p_3$ are respectively the incoming and outgoing momenta of the
external fermions, and momentum conservation requires $q=p_2-p_3$.
The kinematics adopted in this work is the one used in
Ref.~\cite{Sturm:2009kb}:
\begin{equation}
  \label{eq:mom-config}
  p_2^2 = p_3^2 = q^2 = -\mu^2\, .
\end{equation}
Following the convention in the paper above, we denote this symmetric
point by the shorthand ``$\sym$". 
\begin{figure}[!ht]
  \centering
 \includegraphics[width=0.27\textwidth]{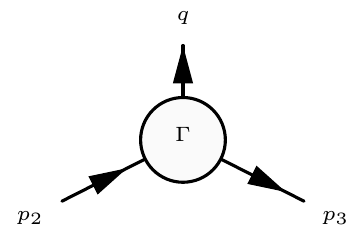}
  \caption{Kinematics used for the correlators of fermion bilinears.}
  \label{fig:kin}
\end{figure}

For the purpose of illustration, we can consider the case of a fermion doublet 
\begin{equation}
  \label{eq:psidoublet}
\psi=\begin{pmatrix}
\psi_1\\
\psi_2
\end{pmatrix}\, ,~~
\overline{\psi}=\begin{pmatrix}
\overline{\psi}_1 & \overline{\psi}_2
\end{pmatrix},
\end{equation}
with mass matrix 
\begin{equation}
  \label{eq:massmatrix}
  \mathcal{M} = \begin{pmatrix}
    m_1 & 0\\
    0 & m_2
  \end{pmatrix} .
\end{equation}
Note that in the mass degenerate case, we simply have
$\mathcal{M}=m\mathds{1}$. If we choose
$\tau^a=\tau^+=\frac{\sigma^+}{2}=\frac{1}{2}\left(\sigma^1+i\sigma^2\right)$, then the
bilinear $O^a_\Gamma=\bar\psi \Gamma \tau^a \psi$ takes the
form $O_\Gamma=\overline{\psi}_1\Gamma{\psi_2}$.\\

The infinitesimal vector and axial non-singlet SU(2) chiral transformation are as follows
\begin{equation}
  \delta\psi(x)=i\Big[\alpha_V(x)\tau^{a}\Big]\psi(x) \, ,~~ 
  \delta\overline{\psi}(x)=-i\overline{\psi}(x)\Big[\alpha_V(x)\tau^{a}\Big],
\end{equation}
and
\begin{equation}
  \delta\psi(x)=i\Big[\alpha_A(x)\tau^{a}\gamma^5\Big]\psi(x) \, , ~~
  \delta\overline{\psi}(x)=i\overline{\psi}(x)\Big[\alpha_A(x)\tau^{a}\gamma^5\Big]
  \, .
\end{equation}

In our conventions, bare quantities are written without any suffix,
while their renormalized counterparts are identified by a suffix
$R$. The renormalization conditions are usually expressed in terms of 
amputated correlators
\begin{equation}
  \label{eq:LambdaO}
  \Lambda^a_\Gamma(p_2,p_3) = S(p_3)^{-1} G^a_\Gamma(p_3,p_2) S(p_2)^{-1}\, ,
\end{equation}
where $S(p)$ is the fermion propagator:
\begin{equation}
  \label{eq:Sdef}
  S(p) = \frac{i}{\pslash - m - \Sigma(p) + i \epsilon}\, .
\end{equation}
Note that for each leg being amputated, the fermion propagator with
the corresponding flavor needs to be used.

The quark mass breaks chiral symmetry explicitly. This breaking is visible in the second equation below, Eq. \ref{eq:AWIbare}. If the regulator does not induce any further breaking of chiral
symmetry, then $\Lambda^a_\gV$ and $\Lambda^a_\gA$ are related to the
fermion propagator by the vector and axial Ward identities
respectively:
\begin{align}
  \label{eq:VWIbare}
  q\cdot \Lambda^a_\gV &= i S(p_2)^{-1} - i S(p_3)^{-1} \, , \\
  \label{eq:AWIbare}
  q\cdot \Lambda^a_\gA &= 2 m i \Lambda^a_\gP - \gamma_5 iS(p_2)^{-1} - iS(p_3)^{-1} \gamma_5\, .
\end{align}
As specified above, the vertex functions are all taken to be non-singlet for the
rest of the paper. In this section the mass-degenerate cases are being
considered, i.e. either both quarks are light (massless) or both are
heavy. In both cases the two fermion propagators that enter in the Ward
identities are the same, and only differ because of the momentum
associated to the external leg. We will suppress the flavor index $a$
to keep the notation simple.

The renormalized quantities are defined as follows: 
\begin{equation}
  \label{eq:Zdef}
  \psi_R = Z_q^{1/2} \psi\, , \quad m_{R} = Z_{m} m\, , \quad M_R =
  Z_M M\, \quad O_{\Gamma,R}
  = Z_\Gamma O_\Gamma\, ,
\end{equation}
where $m$ and $M$ denote the masses of the light and heavy quark
respectively. The renormalized propagator and amputated vertex
functions are 
\begin{equation}
  \label{eq:SGren}
  S_R(p) = Z_q S(p)\, , \quad \Lambda_{\Gamma,R}(p_2,p_3) =
  \frac{Z_\Gamma}{Z_q} \Lambda_\Gamma(p_2,p_3)\, ,
\end{equation}
where $q=l, H$ for light and heavy quarks respectively. Note that our
conventions for defining the fermion propagator are slightly different
from the ones used in Ref.~\cite{Sturm:2009kb}; using our own
conventions, the RI/SMOM conditions are
\begin{align}
  \label{eq:SMOM1}
  \chlim &\left. \frac{1}{12 p^2} \Tr \left[i S_R(p)^{-1}
           \pslash\right] \right|_{p^2=-\mu^2}=1\, ,     \\
  \label{eq:SMOM2}
  \chlim &\frac{1}{12 m_R} \left\{\left. \Tr \left[-i
           S_R(p)^{-1}\right] \right|_{p^2=-\mu^2}
           - \frac12 \left. \Tr \left[\left(q\cdot \Lambda_{\gA,R}\right) \gamma_5
           \right]\right|_\sym \right\}=1\, , \\
  \label{eq:SMOM3}
  \chlim &\frac{1}{12 q^2} \Tr \left. \left[ \left(q
           \cdot \Lambda_{\gV,R} \right) \qslash \right] \right|_\sym
           = 1\, ,\\
  \label{eq:SMOM4}
  \chlim & \frac{1}{12 q^2} \Tr \left. \left[ \left(q
           \cdot \Lambda_{\gA,R} \right) \gamma_5 \qslash \right]
           \right|_\sym = 1\, ,\\
  \label{eq:SMOM5}
  \chlim & \frac{1}{12 i} \Tr \left. \left[ 
           \Lambda_{\gP,R} \gamma_5 \right] \right|_\sym = 1\, ,\\
  \label{eq:SMOM6}
  \chlim & \frac{1}{12} \Tr \left. \left[ 
           \Lambda_{\gS,R} \right] \right|_\sym = 1\, .
\end{align}

These renormalization conditions ensure that the renormalized
bilinears obey vector and axial renormalized Ward identities like the
ones in Eqs.~\ref{eq:VWIbare}, and~\ref{eq:AWIbare}, and the
renormalization constants satisfy the same properties as in the
$\overline{\mathrm{MS}}$ scheme, namely
\begin{equation}
  \label{eq:Zrelations}
  Z_\gV = Z_\gA =1, \quad Z_\gP=Z_\gS, \quad Z_m Z_\gP=1\, .
\end{equation}

While the renormalization conditions in the RI/SMOM scheme are imposed
in the chiral limit, the RI/mSMOM scheme is defined by imposing a
similar set of conditions at some fixed value of a reference
renormalized mass that we denote by $\overline{m}$:
\begin{align}
  \label{eq:mSMOM1}
  \mlim &\left. \frac{1}{12 p^2} \Tr \left[i S_R(p)^{-1}
     \pslash\right] \right|_{p^2=-\mu^2}=1\, ,\\
  \label{eq:mSMOM2}
  \mlim & \frac{1}{12 M_R} \left\{\left. \Tr \left[-i
     S_R(p)^{-1}\right] \right|_{p^2=-\mu^2}
     - \frac12 \left. \Tr \left[\left(q\cdot \Lambda_{\gA,R}\right) \gamma_5
     \right]\right|_\sym \right\}=1\, , \\
   \label{eq:mSMOM3}
  \mlim & \frac{1}{12 q^2} \Tr \left. \left[ \left(q
          \cdot \Lambda_{\gV,R} \right) \qslash \right] \right|_\sym =
          1\, , \\
   \label{eq:mSMOM4}
  \mlim & \frac{1}{12 q^2} \Tr \left. \left[ \left(q
          \cdot \Lambda_{\gA,R} - 2M_R i \Lambda_{\gP, R} \right)
          \gamma_5 \qslash \right] \right|_\sym = 1\, ,\\
   \label{eq:mSMOM5}
  \mlim &\frac{1}{12 i} \Tr \left. \left[ 
          \Lambda_{\gP,R} \gamma_5 \right] \right|_\sym = 1\, ,\\
   \label{eq:mSMOM6}
  \mlim &\Bigg\{\frac{1}{12}\text{Tr}\left[\Lambda_{S,R}\right]-
          \frac{1}{6q^2}\text{Tr}\left[2iM_R\Lambda_{P,R}\gamma_5\slashed{q}\right]\Bigg\}\Bigg|_{\text{sym}}=1\, .
\end{align}
Comparing with the SMOM prescription, only the renormalization
condition for the axial vertex has been modified by a term
proportional to $M_R$, which therefore vanishes in the chiral
limit. We have introduced a new scale $\overline{m}$, which identifies
the renormalized mass at which the renormalization conditions are
imposed. The scale $\overline{m}$ is a free parameter, which needs to be
specified in order to fully define the renormalization scheme. In the
limit where $\overline{m} \to 0$, the mSMOM prescription reduces to
the SMOM one. As usual the renormalization conditions are satisfied by
the tree level values of the field correlators.

The properties of the renormalization constants defined by the mSMOM
conditions can be obtained by following very closely their derivation
in the SMOM schemes. In the case of $Z_\gV$ the derivation is exactly
the same. Using the relation between renormalized and bare vertex
functions, and Eq.~(\ref{eq:mSMOM3}), we obtain
\begin{align}
  \mlim \frac{1}{12 q^2} &\Tr \left. \left[ \left(q
          \cdot \Lambda_{\gV} \right) \qslash \right] \right|_\sym =  
  \mlim \frac{Z_q}{Z_\gV} \frac{1}{12 q^2} \Tr \left. \left[ \left(q
          \cdot \Lambda_{\gV,R} \right) \qslash \right] \right|_\sym
  \\
  \label{eq:cfr1}
  & = \frac{Z_q}{Z_\gV}\, .
\end{align}
Using the vector Ward identity, Eq.~(\ref{eq:VWIbare}), the LHS of the
expression above can be written as
\begin{align}
  \mlim \frac{1}{12 q^2} &\Tr \left. \left[
  \left( i S(p_2)^{-1} - i S(p_3)^{-1}\right) \qslash 
  \right] \right|_\sym = 
  \frac{1}{12 q^2} \Tr \left. \left[
  i S(q)^{-1} \qslash 
  \right] \right|_\sym  \\
  \label{eq:cfr2}
  &= Z_q   \mlim \frac{1}{12 q^2} \Tr \left. \left[
  i S_R(q)^{-1} \qslash 
  \right] \right|_{q^2=-\mu^2} = Z_q  \, .
\end{align}
Comparing Eqs.~(\ref{eq:cfr1}) and~(\ref{eq:cfr2}) yields $Z_\gV=1$.

Because of the modified renormalization condition for the
renormalization of the axial vertex function, the computation of
$Z_\gA$ and $Z_M Z_\gP$ are coupled in the mSMOM scheme. The axial
Ward identity, Eq.~(\ref{eq:AWIbare}), can be rewritten in terms of
renormalized quantities:
\begin{align}
  \label{eq:RWI}
  \frac{1}{Z_\gA} q\cdot \Lambda_{\gA,R} &- \frac{1}{Z_M Z_\gP} 2 M_R
                                           i \Lambda_{\gP,R} = -
                                           \left\{ \gamma_5
                                           iS_R(p_2)^{-1} +
                                           iS_R(p_3)^{-1} \gamma_5 \right\}\, .
\end{align}
Two independent equations can be obtained by multipling
Eq.~(\ref{eq:RWI}) by $\gamma^5 \qslash$ and $\gamma_5$ respectively,
taking the trace, and evaluating correlators at the symmetric
point. In the first case we obtain
\begin{equation}
  \label{eq:ZAeq}
  (Z_\gA -1) = \left(1 - \frac{Z_\gA}{Z_M Z_\gP}\right) C_{mP}\, ,
\end{equation}
where
\begin{equation}
  \label{eq:mPdef}
  C_{m\gP} = \mlim \frac{1}{12 q^2} \Tr \left. \left[2i M_R
      \Lambda_{\gP,R} \gamma_5 \qslash\right]\right|_\sym\, .
\end{equation}
The second equation instead yields
\begin{equation}
  \label{eq:Adel}
  (Z_\gA - 1) C_{qA} = -2 Z_\gA \left(1 - \frac{1}{Z_M Z_\gP}\right)\, ,
\end{equation}
where we have introduced one more constant
\begin{equation}
  \label{eq:mA}Z_\gP
  C_{q\gA} = \mlim \frac{1}{12 M_R} \Tr \left. \left[
      q\cdot \Lambda_{\gA,R} \gamma_5
    \right]\right|_\sym \, .
\end{equation}
It is easy to verify that $Z_\gA=1$, $Z_M Z_\gP=1$ is a solution of the
system. The renormalization constants defined through the mSMOM
prescription do satisfy the properties in Eq.~(\ref{eq:Zrelations}) , as
is the case for the renormalization constants defined in massless
schemes like e.g. RI/SMOM. As a consequence Eq.~(\ref{eq:RWI}) reduces
to the correct axial Ward identity for the renormalized
correlators. Note in particular that $Z_\gA=1$ implies that $Z_\gA$ does
not depend on the renormalization scale $\mu$. As emphasized in
Ref.~\cite{Blum:2001sr}, using the conventional RI/MOM prescription,
these relations are not satisfied in the presence of an explicit
breaking of chiral symmetry. In this respect mSMOM inherits the good properties of the SMOM scheme.

\section{Perturbative computation}
\label{sec:pert-comp}

It is instructive to understand the details of the RI/mSMOM scheme by
performing an explicit one-loop computation. For simplicity we
regularize the theory using dimensional regularization, and evaluate
the relevant diagrams including their dependence on the bare mass
$m$. Because we are mostly interested in flavor non-singlet
quantities, we do not need to worry about extending the definition of
$\gamma_5$ to arbitrary dimensions \cite{'tHooft:1972fi,Breitenlohner:1977hr}. If one were interested in flavor singlet currents, then a
precise definition of $\gamma_5$ in dimensional regulation is
mandatory. In this Section we focus on the actual results, and their
consequences, while we report on the technical details of the
computations in App.~\ref{sec:methods-massive-one}.

\subsection{Fermion self-energy}
\label{sec:fermion-self-energy}

Setting $D=4-2\epsilon$ the fermion self-energy is 
\begin{align}
  \Sigma(p) &= \frac{\alpha}{4\pi} C_2(F)\, \left[ \pslash
              \left(
              -\frac{1}{\epsilon} -1 + \gamma_E + \frac{m^2}{\mu^2} + 
              \frac{m^4}{\mu^4} \ln\left(\frac{m^2}{m^2+\mu^2}\right) 
              + \ln\left(\frac{m^2+\mu^2}{\tilde\mu^2}\right)
              \right) \right. \nonumber \\ 
  \label{eq:SigmaOneLoop}
  & \left. \quad + m \left( 
    \frac{4}{\epsilon} + 6 - 4\gamma_E + \frac{4m^2}{\mu^2}
    \ln\left(\frac{m^2}{m^2+\mu^2}\right)
    - 4 \ln\left(\frac{m^2+\mu^2}{\tilde\mu^2}\right)
    \right) \right]\, ,
\end{align}
where $\gamma_E$ is the Euler-Mascheroni constant, we have
replaced $p^2=-\mu^2$, and denoted $\tilde\mu$ the scale introduced by
dimensional regularization through the rescaling of the gauge coupling
$g \to g\tilde{\mu}^\epsilon$.

Eq.~(\ref{eq:mSMOM1}) yields the renormalization constant for the
fermion field in the mSMOM scheme: 
\begin{equation}
  \label{eq:ZqmSMOM}
  Z_q = 1 + \frac{\alpha}{4\pi} C_2(F) \left[
    \frac{1}{\epsilon} +1 - \gamma_E - \frac{\mb^2}{\mu^2} - 
              \frac{\mb^4}{\mu^4} \ln\left(\frac{\mb^2}{\mb^2+\mu^2}\right) 
              - \ln\left(\frac{\mb^2+\mu^2}{\tilde\mu^2}\right)
    \right]\, .
\end{equation}
The effect of the change of scheme is a redefinition of the finite
part of the renormalization constant $Z_q$. As expected on dimensional
grounds, the dependence on the reference mass $\mb$ only enters via
the dimensionless ratio $\mb/\mu$. The limit for $\mb \to 0$ is well
defined and reproduces the result of the massless scheme~\cite{Sturm:2009kb}.

\subsection{Vector vertex}
\label{sec:vector-vertex}

Let us now start considering the vertex functions, and discuss in
detail the structure of the vector correlator $\Lambda_\gV$. The
one-loop contribution to the vertex for the case of massive fermions
is
\begin{equation}
  \label{eq:LVOneLoop}
  \Lambda^{(1)\sigma}_\gV(p_2,p_3) = -i g^2 C_2(F) \int_k \frac{
    \gamma_\alpha \left[\pslash_3 - \kslash + m\right] \gamma^\sigma 
    \left[\pslash_2 - \kslash + m\right] \gamma^\alpha
  }{
    k^2\left[\left(p_3-k\right)^2-m^2\right] \left[\left(p_2-k\right)^2-m^2\right]
  }\, .
\end{equation}
It is clear from this compact expression that
$\Lambda^{(1)\sigma}_\gV(p_2,p_3)$ transforms as a four-vector under
Lorentz transformations. A closer inspection shows that the integral
can be expressed in terms of just five form factors
\begin{align}
  \label{eq:LVff}
   \Lambda^{(1)\sigma}_\gV(p_2,p_3) = \frac{\alpha}{4\pi} C_2(F) 
  \left[
                                   A_\gV \frac{1}{\mu^2} \right. &
                                                                   \left. \left( i \epsilon^{\sigma \rho \alpha \beta} \gamma_\rho
                                   \gamma^5  p_{3\alpha} p_{2\beta} \right) + 
                                   B_\gV \gamma^\sigma +
                                   C_\gV \frac{1}{\mu^2} \left( p_2^\sigma \pslash_2 +
                                   p_3^\sigma \pslash_3 \right) +  \right.
                                         \nonumber \\
 & \left. + D_\gV \frac{1}{\mu^2} \left( p_2^\sigma \pslash_3 + p_3^\sigma \pslash_2 \right) +
  E_\gV \frac{1}{\mu} \left( p_2^\sigma + p_3^\sigma \right) \right] \, .
\end{align}
The form factors $A_\gV, \ldots, E_\gV$ only depend on the Lorentz
invariants, and are computed analytically. At the symmetric point,
they are given by the following expressions.
\begin{align}
  \label{eq:AV}
  A_\gV = \frac{4}{3} \left[
  \left(\frac12 - \mmu \right) C_0\left(\frac{m^2}{\mu^2}\right) 
  + \left(1+\mmu\right) \log\left(\frac{m^2}{m^2+\mu^2}\right)
  - \sqrt{1 + 4\mmu} 
  \log\left( \frac{\sqrt{1 + 4\mmu} - 1}{\sqrt{1 + 4\mmu} + 1}\right) 
  \right] \, ,
\end{align}
where the expression for $C_0\left(\frac{m^2}{\mu^2}\right)$ can be found in App.~\ref{sec:methods-massive-one}, Eq.~\ref{eq:FinalInt} and Eq.~\ref{eq:numericC0}. Although the last two terms in the expression are separately divergent in
the massless limit, these divergences cancel,
yielding a finite expression when $m \to 0$, which agrees with the
results in Ref.~\cite{Sturm:2009kb}. Similarly for the other form factors we find:

\begin{flalign}
  \label{eq:BV}
  B_\gV = &\frac{1}{\epsilon} - \gamma_E + \frac13 \left[
  - C_0\left(\frac{m^2}{\mu^2}\right) \left( 1 - 4 \mmu -2 \mmmu \right) 
  + 2 \left( 3 - \mmu \right) \mmu \plog{\frac{m^2}{m^2 + \mu^2}}
            + \left( 1 - 4\mmu \right) \plog{\frac{m^2}{\tilde\mu^2}}
            \right.
            \nonumber \\
  & \left. 
    - 4 \left( 1 - \mmu \right) \plog{\frac{m^2+\mu^2}{\tilde\mu^2}}
    - \left( 1- 2\mmu \right) \sqrt{1+4\mmu} 
  \log\left( \frac{\sqrt{1 + 4\mmu} - 1}{\sqrt{1 + 4\mmu} + 1}\right) 
  \right]\, ; &&
\end{flalign}

\begin{flalign}
  \label{eq:CV}
  C_\gV = -\frac{2}{3} & \left[
                         \left( 1 - \mmu \right) \mmu
                         \plog{\frac{m^2}{m^2+\mu^2}} +
                         \left( 1 -2 \mmu \right) \sqrt{1+4\mmu} 
  \log\left( \frac{\sqrt{1 + 4\mmu} - 1}{\sqrt{1 + 4\mmu} + 1}\right) 
                         \right. \nonumber \\
  &\left. 
  + \left( 2 - \mmu \right) - 2 C_0\left(\frac{m^2}{\mu^2}\right) \mmu \left( 1 + \mmu \right)
    - \left( 1 - 4\mmu \right) \plog{\frac{m^2}{\tilde\mu^2}} 
    + \left( 1 - 4\mmu \right) \plog{\frac{m^2+\mu^2}{\tilde\mu^2}} 
  \right] \, ; &&
  \end{flalign}

\begin{align}
  \label{eq:DV}
  D_\gV = \frac{2}{3} &\left[
  \left( 1+ C_0\left(\frac{m^2}{\mu^2}\right) \right) \left( 1 - 2\mmu \right) -2 \left( 1 + \mmu
  \right) \mmu \plog{\frac{m^2}{m^2+\mu^2}} 
  \right]\, ;  \ \ \ \ \ \ \ \ \ \ \ \ \ \ \ \ \ \ \ \ \ \ \ \ \ \ \ \ \ \ \ \ \ \ \ \ \ \ \ \ \ \ \ \ \ \ \ \ \ \ \ \ 
\end{align}

\begin{flalign}
  \label{eq:EV}
  E_\gV = -\frac{4}{3} & \frac{m}{\mu} \Bigg[
  C_0\left(\frac{m^2}{\mu^2}\right) \left( 1 - 2\mmu \right) + 2 \plog{\frac{m^2}{m^2+\mu^2}}
  + 2 \mmu \plog{\frac{m^2}{m^2+\mu^2}} 
\nonumber \\ 
& \ \ \ \ 
 -2 \sqrt{1+4\mmu} 
  \log\left( \frac{\sqrt{1 + 4\mmu} - 1}{\sqrt{1 + 4\mmu} + 1}\right) 
  \Bigg]\, ; &&
  \end{flalign}
which all agree with the results in Ref.~\cite{Sturm:2009kb} when the limit $m\to 0$ is taken.

\subsection{Pseudoscalar vertex}
\label{sec:pseudoscalar-vertex}
For the pseudoscalar vertex function at one-loop we have:

\begin{equation}
  \label{eq:LPOneLoop}
  \Lambda^{(1)}_\gP(p_2,p_3) = g^2 C_2(F) \int_k \frac{
    \gamma_\alpha \left[\pslash_3 - \kslash + m\right] \gamma^5
    \left[\pslash_2 - \kslash + m\right] \gamma^\alpha
  }{
    k^2\left[\left(p_3-k\right)^2-m^2\right] \left[\left(p_2-k\right)^2-m^2\right]
  }\, .
\end{equation}
The one-loop structure of this vertex is simpler
\begin{align}
  \label{eq:LPff}
   \Lambda^{(1)}_\gP(p_2,p_3) = \frac{i\alpha}{4\pi} C_2(F) 
  \left[
                                   B_\gP \left( \gamma^5 \right) 
  + E_\gP \frac{1}{\mu} (\gamma^5) \left( \pslash_2 - \pslash_3 \right) \right] \, .
\end{align}
The form factors are:
\begin{align}
  \label{eq:BP}
  B_\gP = 4 \left[
  \frac{1}{\epsilon} - \gamma_E + \frac{3}{2}  - \frac12 C_0\left(\frac{m^2}{\mu^2}\right) 
  + \mmu \plog{\frac{m^2}{m^2+\mu^2}} - \plog{\frac{m^2+\mu^2}{\tilde{\mu}^2}}
  \right]\, ;
\end{align}

\begin{align}
  \label{eq:EP}
  E_\gP = -\frac{m}{\mu} 2C_0\left(\frac{m^2}{\mu^2}\right)\, .
\end{align}

Using the renormalization condition Eq.~(\ref{eq:mSMOM5}), we have

\begin{align}
 \mlim &\frac{1}{12 i} \Tr \left. \left[ 
          \Lambda_{\gP,R} \gamma_5 \right] \right|_\sym =\lim_{m_R\to
         \overline{m}}
         \frac{1}{12i}\text{Tr}\left[\frac{Z_\gP}{Z_q}\Lambda_{P}\gamma^5\right]\Bigg|_{\text{sym}}= 1,
\end{align}

giving
\begin{align}
Z_\gP=\Bigg\{1+
  \frac{\alpha}{4\pi} C_2(F) \Bigg[ & -3
  \left(\frac{1}{\epsilon}-\gamma_E\right) - 5 + 2C_0\left(\frac{m^2}{\mu^2}\right) -
  \frac{\overline{m}^2}{\mu^2}\left( 1 - 4 \ln\left( 1 +
  \frac{\mu^2}{\overline{m}^2} \right)
  -\frac{\overline{m}^2}{\mu^2} \ln\left( 1 +
  \frac{\mu^2}{\overline{m}^2} \right) \right)  \nonumber \\
  & + 3\ln\left(\frac{\overline{m}^2+\mu^2}{\tilde{\mu}^2}\right)\Bigg]\Bigg\}.
\end{align}

The above result reduce to Ref.~\cite{Sturm:2009kb} in the massless limit. Note that $Z_\gP$ is scale dependent; setting $\tilde\mu=\mu$, we find
that the dependence on the scale only appears through the combination $\mu/\overline{m}$.

\subsection{Axial vertex}
\label{sec:axial-vertex}

The computation of the axial vertex follows very closely the one of
the vector vertex presented above. The starting expression
\begin{align}
  \label{eq:LAOneLoop}
  \Lambda^{(1)\sigma}_\gA(p_2,p_3) = -i g^2 C_2(F) \int_k \frac{
    \gamma_\alpha \left[\pslash_3 - \kslash + m\right] \gamma^\sigma \gamma^5
    \left[\pslash_2 - \kslash + m\right] \gamma^\alpha
  }{
    k^2\left[\left(p_3-k\right)^2-m^2\right] \left[\left(p_2-k\right)^2-m^2\right]
  }  
\end{align}
can again be parametrized in terms of five form factors, which we
denote $A_\gA, \ldots, E_\gA$, 
\begin{align}
  \label{eq:LAff}
  \Lambda^{(1)\sigma}_\gA(p_2,p_3) = \frac{\alpha}{4\pi} C_2(F) 
  \left[
                                   A_\gA \frac{1}{\mu^2} \right. &
                                                                   \left. \left( i \epsilon^{\sigma \rho \alpha \beta} \gamma_\rho
                                   p_{3\alpha} p_{2\beta} \right) + 
                                   B_\gA \gamma^\sigma \gamma^5 +
                                   C_\gA \frac{1}{\mu^2} \gamma^5 \left( p_2^\sigma \pslash_2 +
                                   p_3^\sigma \pslash_3 \right) + \right.
                                         \nonumber \\
 & \left. + D_\gA \frac{1}{\mu^2} \gamma^5 \left( p_2^\sigma \pslash_3 + p_3^\sigma \pslash_2 \right) +
  E_\gA \frac{1}{\mu} \left( p_2^\sigma - p_3^\sigma \right) \right] \, .
\end{align}
For the axial form factors we find: 
\begin{align}
  \label{eq:AA}
  A_\gA = \frac{4}{3} \left[
  \left(\frac12 - \mmu \right) C_0\left(\frac{m^2}{\mu^2}\right) 
  + \mmu \log\left(\frac{m^2}{m^2+\mu^2}\right)
  - \plog{\frac{m^2+\mu^2}{\tilde\mu^2}}
  - \sqrt{1 + 4\mmu} 
  \log\left( \frac{\sqrt{1 + 4\mmu} - 1}{\sqrt{1 + 4\mmu} + 1}\right) 
  \right] \, ;
\end{align}

\begin{align}
  \label{eq:BA}
  B_\gA = &\frac{1}{\epsilon} - \gamma_E + \frac13 \left[
  - C_0\left(\frac{m^2}{\mu^2}\right) \left( 1 + 8 \mmu - 2 \mmmu \right) 
  + \left( 3 - \mmu \right) 2 \mmu \plog{\frac{m^2}{m^2 + \mu^2}}
            + \left( 1 - 4\mmu \right) \plog{\frac{m^2}{\tilde\mu^2}}
            \right.
            \nonumber \\
  & \left. 
    - 4 \left( 1 - \mmu \right) \plog{\frac{m^2+\mu^2}{\tilde\mu^2}}
    - \left( 1- 2\mmu \right) \sqrt{1+4\mmu} 
  \log\left( \frac{\sqrt{1 + 4\mmu} - 1}{\sqrt{1 + 4\mmu} + 1}\right) 
  \right]\, ;
\end{align}

\begin{align}
  \label{eq:CA}
  C_\gA = -\frac{2}{3} & \left[
                         \left( 4 - \mmu \right) \mmu
                         \plog{\frac{m^2}{m^2+\mu^2}} -
                         \left( 1 -2 \mmu \right) \sqrt{1+4\mmu} 
  \log\left( \frac{\sqrt{1 + 4\mmu} - 1}{\sqrt{1 + 4\mmu} + 1}\right) 
                         \right. \nonumber \\
  &\left. 
  - \left( 2 - \mmu \right) + 2 C_0\left(\frac{m^2}{\mu^2}\right) \mmu \left( 1 + \mmu \right)
    + \left( 1 - 4\mmu \right) \plog{\frac{m^2}{\tilde\mu^2}} 
    - \left( 1 - 4\mmu \right) \plog{\frac{m^2+\mu^2}{\tilde\mu^2}} 
  \right] \, ;
\end{align}

\begin{align}
  \label{eq:DA}
  D_\gA = -\frac{2}{3} \left[
  \left( 1+ C_0\left(\frac{m^2}{\mu^2}\right) \right) \left( 1 - 2\mmu \right) -2 \left( 1 + \mmu
  \right) \mmu \plog{\frac{m^2}{m^2+\mu^2}} 
  \right]\, ;
\end{align}

\begin{align}
  \label{eq:EA}
  E_\gA = \frac{m}{\mu} 4C_0\left(\frac{m^2}{\mu^2}\right)\, .
\end{align}

Again, in the massless limit $m\to 0$, the above coefficients coincide
with the corresponding results in Ref.~\cite{Sturm:2009kb}.

\subsection{Scalar vertex}
\label{sec:scalar-vertex}

In this section we discuss the mSMOM renormalization condition for the scalar vertex. 
\begin{equation}
  \label{eq:LSOneLoop}
  \Lambda^{(1)}_\gS(p_2,p_3) = -ig^2 C_2(F) \int_k \frac{
    \gamma_\alpha \left[\pslash_3 - \kslash + m\right]
    \left[\pslash_2 - \kslash + m\right] \gamma^\alpha
  }{
    k^2\left[\left(p_3-k\right)^2-m^2\right] \left[\left(p_2-k\right)^2-m^2\right]
  }\, .
\end{equation}
The one-loop structure of this vertex is 
\begin{align}
  \label{eq:LPff}
   \Lambda^{(1)}_\gS(p_2,p_3) = \frac{\alpha}{4\pi} C_2(F) 
  \left[
                                   B_\gS 
  + E_\gS \frac{1}{\mu}  \left( \pslash_2 + \pslash_3 \right) \right] \, .
\end{align}
The form factors are:

\begin{align}
  \label{eq:BS}
  B_\gS =\Bigg\{4\left(\frac{1}{\epsilon}-\gamma_E\right)+6-\left(8\frac{m^2}{\mu^2}+2\right)C_0\left(\frac{m^2}{\mu^2}\right)+\frac{4m^2}{\mu^2}\ln\left(\frac{m^2}{m^2+\mu^2}\right)-4\ln\left(\frac{m^2+\mu^2}{\tilde{\mu}^2}\right) ,
\end{align}

\begin{align}
  \label{eq:ES}
  E_\gS = -\frac{4}{3} & \frac{m}{\mu} \left[
  C_0\left(\frac{m^2}{\mu^2}\right) \left( -\frac{1}{2} + \mmu \right) -\left(1+\frac{m^2}{\mu^2}\right) \plog{\frac{m^2}{m^2+\mu^2}}
 + \sqrt{1+4\mmu} 
  \log\left( \frac{\sqrt{1 + 4\mmu} - 1}{\sqrt{1 + 4\mmu} + 1}\right) 
  \right]\, .
\end{align}

Using the renormalization condition Eq.~(\ref{eq:mSMOM6}), and the fact that $Z_mZ_\gP=1$, yields
\begin{equation}
\begin{split}
&\lim_{m_R\to \overline{m}}\Bigg\{\frac{1}{12}\text{Tr}\left[\frac{Z_S}{Z_q}\Lambda_{S}\right]+\frac{1}{6q^2}\text{Tr}\left[\frac{Z_mZ_\gP}{Z_q}2im\Lambda_{P}\gamma_5\slashed{q}\right]\Bigg\}\Bigg|_{\text{sym}}\\=
&\lim_{m_R\to \overline{m}}Z_q^{-1}\Bigg\{ Z_S\Bigg(1+C_2(F)\frac{\alpha}{4\pi}\Bigg[4\left(\frac{1}{\epsilon}-\gamma_E\right)+6-\left(8\frac{m^2}{\mu^2}+2\right)C_0\left(\frac{m^2}{\mu^2}\right)+\frac{4m^2}{\mu^2}\ln\left(\frac{m^2}{m^2+\mu^2}\right)-4\ln\left(\frac{m^2+\mu^2}{\tilde{\mu}^2}\right)\Bigg)\\
&\ \ \ \ \ \ \ \ \ \ \ \ \ \ \ +\frac{8m^2}{\mu^2}C_0\left(\frac{m^2}{\mu^2}\right)\Bigg]\Bigg\}=1\, .
\end{split}
\end{equation}
After introducing 
\begin{equation}
\mathcal{P}=\Bigg(1+C_2(F)\frac{\alpha}{4\pi}\Bigg[4\left(\frac{1}{\epsilon}-\gamma_E\right)+6-2C_0\left(\frac{m^2}{\mu^2}\right)+\frac{4m^2}{\mu^2}\ln\left(\frac{m^2}{m^2+\mu^2}\right)-4\ln\left(\frac{m^2+\mu^2}{\tilde{\mu}^2}\right)\Bigg)\, ,
\end{equation}
we obtain
\begin{equation}
\begin{split}
Z_S\Bigg(\mathcal{P}-\frac{\alpha}{4\pi}C_2(F)\frac{8m^2}{\mu^2}C_0\left(\frac{m^2}{\mu^2}\right)\Bigg)&=Z_q\left(1-\frac{1}{Z_q}C_2(F)\frac{\alpha}{4\pi}\frac{8m^2}{\mu^2}\right)
\nonumber \\ 
&=Z_q\left(1-C_2(F)\frac{\alpha}{4\pi}\frac{8m^2}{\mu^2}+\mathcal{O}(\alpha^2)\right)\, ,
\end{split}
\end{equation}
and hence
\begin{equation}
\begin{split}
Z_S=&Z_q\mathcal{P}^{-1}\left(1-C_2(F)\frac{\alpha}{4\pi}\frac{8m^2}{\mu^2}+\mathcal{O}(\alpha^2)\right)\Bigg(1+\frac{\alpha}{4\pi}C_2(F)\frac{8m^2}{\mu^2}\frac{C_0\left(\frac{m^2}{\mu^2}\right)}{\mathcal{P}}\Bigg)\\=
&Z_q\mathcal{P}^{-1}\left(1-C_2(F)\frac{\alpha}{4\pi}\frac{8m^2}{\mu^2}+\mathcal{O}(\alpha^2)\right)\Bigg(1+\frac{\alpha}{4\pi}C_2(F)\frac{8m^2}{\mu^2}C_0\left(\frac{m^2}{\mu^2}\right)+\mathcal{O}(\alpha^2)\Bigg)\\=
&Z_\gP \,.
\end{split}
\end{equation}

We can rewrite the above expression explicitly as:
\begin{equation}
\begin{split}
Z_S=&\Bigg\{1+ \frac{\alpha}{4\pi} C_2(F) \Bigg[ -3
  \left(\frac{1}{\epsilon}-\gamma_E\right) - 5 + 2C_0\left(\frac{m^2}{\mu^2}\right) \\
  & \ \ \ \ \ \ \ \ \ \ \ \ \ \ \ \ \ \ \ \ \ -
  \frac{\overline{m}^2}{\mu^2}\left( 1 - 4 \ln\left( 1 +
  \frac{\mu^2}{\overline{m}^2} \right) 
  -\frac{\overline{m}^2}{\mu^2} \ln\left( 1 +
  \frac{\mu^2}{\overline{m}^2} \right) \right) + 3\ln\left(\frac{\overline{m}^2+\mu^2}{\tilde{\mu}^2}\right)\Bigg]\Bigg\} \\ =
 & Z_\gP
 \end{split}
\end{equation}
which clearly depends on the ratio $\frac{m^2}{\mu^2}$. 

It is possible to show non-perturbatively that $Z_mZ_{S}=1$ using the vector WI
with a suitable probe. See \eg Ref.~\cite{Vladikas:2011bp} for a
detailed discussion.

\subsection{Mass Renormalization}
The mass renormalization can be computed following the mSMOM prescription:

\begin{equation}
\lim_{m_R\to \overline{m}}\frac{1}{12m_R}\left\{\text{Tr}\Bigg[-iS^{-1}_R\Bigg]-\frac{1}{2}\text{Tr}\Bigg[q_\mu\Lambda_{A,R}^\mu\gamma^5\Bigg]\right\}\Bigg|_{\text{sym}}=1.
\end{equation}
We prove that $Z_mZ_\gP$ has to be equal to 1, i.e. 
\begin{equation}
\begin{split}
&\lim_{m_R\to \overline{m}}\frac{1}{12Z_m m}\left\{\text{Tr}\Bigg[-iZ_q^{-1}S^{-1}\Bigg]-\frac{1}{2}\text{Tr}\Bigg[Z_\gA Z_q^{-1}q_\mu\Lambda_{A,R}^\mu\gamma^5\Bigg]\right\}\Bigg|_{\text{sym}}\\=
&\lim_{m_R\to \overline{m}}\frac{Z_m^{-1}}{12m}\left\{Z_q^{-1}(12m)(1+\Sigma_S(p^2))-\frac{1}{2}Z_\gA Z_q^{-1}(12)C_2(F)\frac{\alpha}{4\pi}4mC_0\left(\frac{m^2}{\mu^2}\right)\right\}\Bigg|_{\text{sym}}
\end{split}
\end{equation}
Setting $Z_\gA=1$, we have
\begin{equation}
\begin{split}
Z_m=&Z_q^{-1}\Bigg[1+\frac{\alpha}{4\pi}C_2(F)\Bigg(4\left(\frac{1}{\epsilon}-\gamma_E\right)+6+\frac{4m^2}{\mu^2}\ln\left(\frac{m^2}{m^2+\mu^2}\right)-4\ln\left(\frac{m^2+\mu^2}{\tilde{\mu}^2}\right)-2C_0\left(\frac{m^2}{\mu^2}\right)\Bigg)\Bigg]\\=
&1+\frac{\alpha}{4\pi} C_2(F) \Bigg[ 3
  \left(\frac{1}{\epsilon}-\gamma_E\right) + 5 - 2C_0\left(\frac{m^2}{\mu^2}\right)\\ 
&\ \ \ \ \ \ \ \ \ \ \ \ \ \ \ \ \ \ + \frac{\overline{m}^2}{\mu^2}\left( 1 + 4 \ln\left(\frac{\overline{m}^2}{\overline{m}^2+\mu^2} \right)
  -\frac{\overline{m}^2}{\mu^2} \ln\left( \frac{\overline{m}^2}{\overline{m}^2+\mu^2} \right) \right)- 3\ln\left(\frac{\overline{m}^2+\mu^2}{\tilde{\mu}^2}\right)\Bigg]\\=
& Z_\gP^{-1}\, .
\end{split}
\end{equation}

\subsection{Vector Ward identity}
\label{sec:vector-ward-identity}

The results in the two previous subsections need to satisfy 
the vector Ward identity. This requirement provides a stringent test of our computations. At one-loop the Ward identity becomes
\begin{align}
  \label{eq:OneLoopVWI}
  q \cdot \Lambda_\gV^{(1)} = \Sigma(p_3) - \Sigma(p_2)\, .
\end{align}
Using the results in Sec.~(\ref{sec:vector-vertex}), the LHS of
Eq.~\ref{eq:OneLoopVWI} is readily evaluated
\begin{align}
  \label{eq:OneLoopVWILHS}
  \frac{\alpha}{4\pi} C_2(F) \qslash \left\{
  \frac{1}{\epsilon} - \gamma_E + 1 -
  \plog{\frac{m^2+\mu^2}{\tilde\mu^2}} -
  \mmu \left(
  1 - \mmu \left[ 1 - \mmu \plog{\frac{m^2}{m^2+\mu^2}} \right]
  \right)
  \right\}\, .
\end{align}
Likewise, for the RHS of Eq.~(\ref{eq:OneLoopVWI}), the results in
Sec.~(\ref{sec:fermion-self-energy}) yield exactly the same expression,
so that the vector Ward identity is indeed satisfied. 

As discussed in the previous section, the vector Ward identity implies
that $Z_V=1$. This can be checked explicitly from our one-loop
calculation. Using the renormalization condition
Eq.~(\ref{eq:mSMOM3}) yields

\begin{align}
 \mlim & \frac{1}{12 q^2} \Tr \left. \left[ \left(q
          \cdot \Lambda_{\gV,R} \right) \qslash \right] \right|_\sym =\lim_{m_R\to \overline{m}}\frac{1}{12q^2}\text{Tr}\left[\frac{Z_V}{Z_q}(q\cdot\Lambda_{V})\qslash\right]\Bigg|_{\text{sym}}=1,
\end{align}
which, using Eq.~(\ref{eq:ZqmSMOM}), implies
\begin{align}
Z_\gV=Z_q\left[1+\frac{\alpha}{4\pi}C_2(F)\Bigg( \frac{1}{\epsilon} +1 - \gamma_E - \frac{\mb^2}{\mu^2} - 
              \frac{\mb^4}{\mu^4} \ln\left(\frac{\mb^2}{\mb^2+\mu^2}\right) 
              - \ln\left(\frac{\mb^2+\mu^2}{\tilde\mu^2}\right)\Bigg)\right]^{-1}=1.
\end{align}

\subsection{Axial Ward identity}
\label{sec:axial-ward-identity}

The axial Ward identity also needs to be fulfilled in our check at 1-loop. This constraint becomes
\begin{align}
\label{eq:OneLoopAWI}
  q \cdot \Lambda_\gA^{(1)}=2mi\Lambda_\gP+\gamma_5\Sigma(p_2)+\Sigma(p_3)\gamma_5
\end{align}
Using the results in Sec.~(\ref{sec:axial-vertex}), the LHS of Eq.~(\ref{eq:OneLoopAWI}) can be evaluated 
\begin{align}
-\frac{\alpha}{4\pi}C_2(F)\gamma^5\Bigg\{\slashed{q}\Bigg[\frac{1}{\epsilon}-\gamma_E+1-\frac{4m^2}{\mu^2}C_0\left(\frac{m^2}{\mu^2}\right)-\frac{m^2}{\mu^2}-\frac{m^4}{\mu^4}\ln\left(\frac{m^2}{m^2+\mu^2}\right)-\ln\left(\frac{m^2+\mu^2}{\tilde{\mu}^2}\right)\Bigg] -4mC_0\left(\frac{m^2}{\mu^2}\right)\Bigg\}.
\end{align}
Similarly, for the RHS of Eq.~(\ref{eq:OneLoopAWI}) , the results in
Sec.~(\ref{sec:fermion-self-energy}) and Sec.~(\ref{sec:pseudoscalar-vertex}) yield exactly the same expression, so that the axial Ward identity is indeed satisfied. 

As discussed in the previous section, the axial Ward identity implies that $Z_\gA=1$. This can be checked explicitly from our one-loop calculation. Note that the modified renormalization condition Eq.~(\ref{eq:mSMOM4}) is critical to get $Z_\gA=1$. 
\begin{align}
 & \mlim  \frac{1}{12 q^2} \Tr \left. \left[ \left(q
          \cdot \Lambda_{\gA,R} - 2m_R i \Lambda_{\gP, R} \right)
          \gamma_5 \qslash \right] \right|_\sym =\mlim\frac{1}{12q^2}\text{Tr}\left[\left(\frac{Z_\gA}{Z_q}q\cdot\Lambda_{A}-\frac{Z_\gP Z_m}{Z_q}2im\Lambda_{P}\right)\gamma^5\slashed{q}\right]\Bigg|_{\text{sym}} \\ \nonumber=
&        \mlim \frac{1}{12q^2}\frac{1}{Z_q}  \text{Tr}\Bigg\{Z_\gA\Bigg(q^2+\frac{\alpha}{4\pi}C_2(F)q^2\Bigg[\frac{1}{\epsilon}-\gamma_E+1-\frac{4m^2}{\mu^2}C_0\left(\frac{m^2}{\mu^2}\right)-\frac{m^2}{\mu^2}\\ \nonumber
& \ \ \ \ \ \ \ \ \ \ \ \ \ \ \ \ \ \ \ \ \ \ \ \ \ \ \ \ \ \ \ \ \ \ \ \ \ \ \ \ \ \ \ \ \ \ \ \ \ \ \ \ \ -\frac{m^4}{\mu^4}\ln\left(\frac{m^2}{m^2+\mu^2}\right)-\ln\left(\frac{m^2+\mu^2}{\tilde{\mu}^2}\right)\Bigg]\Bigg)\\ \nonumber
& \ \ \ \ \ \ \ \ \ \ \ \ \ \ \ \ \ \ \ \ \ \ \ \ +C_2(F)\frac{\alpha}{4\pi}q^2\frac{4m^2}{\mu^2}C_0\left(\frac{m^2}{\mu^2}\right)\Bigg\}\Bigg|_{\text{sym}}=1,
\end{align}
where we have used $Z_mZ_\gP=1$. Substituting Eq.~(\ref{eq:ZqmSMOM}), yields
\begin{align}
Z_\gA=1 \, .
\end{align}

\section{Mass non-degenerate scheme}
\label{sec:mass-non-degenerate-scheme}
We will now consider the renormalization scheme for the case of
non-singlet, mass non-degenerate vertex functions. Note that according
to Eq.~\ref{eq:psidoublet}, we collect the two fermion fields in a
flavor doublet:
\begin{equation}
  \label{eq:psidoublet-mixed}
  \psi=\begin{pmatrix}
    H\\
    l
  \end{pmatrix} \,  , ~~
  \overline{\psi}=\begin{pmatrix}
    \overline{H} & \overline{l}
  \end{pmatrix},
\end{equation}
with the non-degenerate mass matrix 
\begin{equation}
  \label{eq:massmatrix-mixed}
  \mathcal{M} = \begin{pmatrix}
    M & 0\\
    0 & m
  \end{pmatrix} .
\end{equation}
In what follows we will be interested in fermion bilinears of the form
$O^+=\overline{H} \Gamma l$ by choosing the flavor rotation matrix to be $\tau^a=\tau^+=\frac{\sigma^+}{2}=\frac{1}{2}\left(\sigma^1+i\sigma^2\right)$. For clarity, we will leave the flavor
index $``+"$ explicit in the Ward identities, but will suppress it for the
rest of the section to keep the notation simple. We have used curly
letters ($\mathcal{V}, \mathcal{A}, \mathcal{P}, \mathcal{S}$) to
denote the heavy-light bilinears. The vector and axial Ward identities
are as follows:
\begin{align}
  \label{eq:MixedVecWI}
q \cdot \Lambda^+_{\mathcal{V}}=(M-m) \Lambda^+_{\mathcal{S}}+iS_{H}(p_2)^{-1}-iS_{l}(p_3)^{-1}.
\end{align}

\begin{align} 
 \label{eq:m=MixedAxialWI}
q\cdot \Lambda^+_\mathcal{A} = (M+m) i \Lambda^+_\mathcal{P} - \gamma_5 iS_H(p_2)^{-1} - iS_l(p_3)^{-1} \gamma_5\, ,
\end{align}
where $M$ and $m$ are masses of the heavy and the light quarks
respectively. 

\subsection{Modified renormalization conditions}
\label{sec:modif-renorm-cond}

The RI/mSMOM scheme for the heavy-light mixed case is defined by
imposing the following set of conditions at some reference mass
$\overline{m}$:
\begin{align}
  \label{eq:mixedmSMOM3}
  \lim_{\substack{m_R\to0 \\ M_R\to\overline{m}}} & \frac{1}{12 q^2} \Tr \left. \left[ \left(q
                                               \cdot \Lambda_{\mathcal{V},R} -(M_R-m_R)\Lambda_{\mathcal{S},R} \right) \qslash \right] \right|_\sym =
                                               \lim_{\substack{m_R\to0 \\ M_R\to\overline{m}}}  \frac{1}{12 q^2} \Tr  \left[ \left(i\zeta^{-1}S_{H,R}(p_2)^{-1}-i\zeta S_{l,R}(p_3)^{-1}\right)
  \qslash \right] 
  \, , \\
  \label{eq:mixedmSMOM4}
  \lim_{\substack{m_R\to0 \\ M_R\to\overline{m}}}  & \frac{1}{12 q^2} \Tr \left. \left[ \left(q
                                                \cdot \Lambda_{\mathcal{A},R}- (M_R+m_R) i \Lambda_{\mathcal{P}, R}\right)
                                                \gamma_5 \qslash \right] \right|_\sym = \nonumber \\
&   \ \ \ \ \ \ \ \ \ \ \ \ \ \ \ \ \ \ \ \ \ \ \ \ \ \ \ \ \ \ \ \ \ \ \ \ \ \ \ \ \ \ \ \ \ \ \ \ \ \ \ \ \ \lim_{\substack{m_R\to0 \\ M_R\to\overline{m}}}    \frac{1}{12 q^2} \Tr  \big[ \big(
  -i\gamma^5\zeta^{-1}S_{H,R}(p_2)^{-1} -i\zeta S_{l,R}(p_3)^{-1}\gamma^5 \big)
  \gamma_5 \qslash \big] \, ,\\
  \label{eq:mixedmSMOM5}
  \lim_{\substack{m_R\to0 \\ M_R\to\overline{m}}}   &\frac{1}{12 i} \Tr \left. \left[ 
                                                 \Lambda_{\mathcal{P},R} \gamma_5 \right] \right|_\sym 
                                                 = \lim_{\substack{m_R\to0 \\ M_R\to\overline{m}}}  \Bigg\{
  \frac{1}{12(M_R+m_R)}\left\{\left. \Tr \left[-i
  \zeta^{-1} S_{H,R}(p)^{-1}\right] \right|_{p^2=-\mu^2}
  - \frac12 \left. \Tr \left[\left(q\cdot \Lambda_{\mathcal{A},R}\right) \gamma_5
  \right]\right|_\sym \right\}+  \nonumber \\
                                             & \ \ \ \ \ \ \ \ \ \ \ \ \ \ \ \ \ \ \ \ \ \ \ \ \ \ \ \ \ \ \ \ \ \ \ \ \ \ \ \  \frac{1}{12(M_R+m_R)}\left\{\left. \Tr \left[-i
                                               \zeta S_{l,R}(p)^{-1}\right] \right|_{p^2=-\mu^2}
                                               - \frac12 \left. \Tr \left[\left(q\cdot \Lambda_{\mathcal{A},R}\right) \gamma_5
                                               \right]\right|_\sym \right\}
                                               \Bigg\}\, .
\end{align}
where $\zeta$ denotes the ratio of the light to the heavy field
renormalizations, i.e. $\zeta=\frac{\sqrt{Z_l}}{\sqrt{Z_H}}$. In the
degenerate mass, $\zeta=1$ and the mixed mSMOM prescription reduces
to the mSMOM and SMOM one. Note that $M$ refers to the heavy quark
mass while the light quark is denoted by $m$ and curly subscripts
denote heavy-light mixed vertices. The renormalization conditions for
$Z_l, Z_H$ and $Z_m$ remain unaltered as they are independently
determined from the corresponding degenerate, massive and massless schemes of the previous sections. As usual the
renormalization conditions are satisfied by the tree level values of
the field correlators.

\subsection{Renormalization constants}
\label{sec:renorm-const}

The properties of the renormalization constants in this scheme are
obtained once again from the Ward identities. We multiply the vector
Ward identity Eq.~\ref{eq:MixedVecWI} by $\slashed{q}$, take the trace and
write the bare quantities in terms of the renormalized ones as
follows:
\begin{align}
  Z_H^{1/2}Z_l^{1/2}  \Tr  \left[ \frac{1}{Z_{\mathcal{V}}}\left(q
  \cdot \Lambda_{\mathcal{V},R} \right) \qslash \right] =
  Z_H^{1/2}Z_l^{1/2}\Tr  \left[ \left(i\zeta^{-1}S_{H,R}(p_2)^{-1}-
  i\zeta
  S_{l,R}(p_3)^{-1}+\frac{\frac{M_R}{Z_M}-\frac{m_R}{Z_m}}{Z_\mathcal{S}}\Lambda_{\mathcal{S},R}\right)
  \qslash \right] \, .
\end{align}
Using Eq.~\ref{eq:mixedmSMOM3} we get
\begin{align}
\left(\frac{1}{Z_\mathcal{V}}-1\right)\Tr  \left[ \left(i\zeta^{-1}S_{H,R}(p_2)^{-1}-i\zeta S_{l,R}(p_3)^{-1}\right) \qslash \right]=\left(\frac{-(M_R-m_R)}{Z_\mathcal{V}}+\frac{\frac{M_R}{Z_M}-\frac{m_R}{Z_m}}{Z_\mathcal{S}}\right)
\Tr  \left[ \Lambda_{\mathcal{S},R} \qslash \right]\, ,
\end{align}
which has a solution when $Z_{\mathcal{V}}=1$ and 
\begin{align}
Z_\mathcal{S}=\frac{\frac{M_R}{Z_M}-\frac{m_R}{Z_m}}{M_R-m_R} \, .
\end{align}

For the axial current we follow a similar procedure, starting from
the bare mixed axial Ward identity
Eq.~\ref{eq:m=MixedAxialWI}. Multiplying once by $\gamma^5 \qslash$
and $\gamma_5$ respectively and taking the trace gives two independent
equations. In the first case, we use Eq.~\ref{eq:mixedmSMOM4} and
obtain
\begin{align}
  \label{eq:mixedaxialgamma5slashq}
(1-\frac{1}{Z_{\mathcal{A}}})\Tr  \left[
  \left(-i\gamma^5\zeta^{-1}S_{H,R}(p_2)^{-1}-i\zeta
  S_{l,R}(p_3)^{-1}\gamma^5\right) \gamma^5\qslash
  \right]=&\left(\frac{M_R+m_R}{Z_{\mathcal{A}}}-\left(\frac{M_R}{Z_MZ_\mathcal{P}}+\frac{m_R}{Z_mZ_\mathcal{P}}\right)\right)Tr  \left[ \left(i\Lambda_{\mathcal{P}}\right) \gamma^5\qslash
  \right]\, .
\end{align}
The latter equation is satisfied by $Z_{\mathcal{A}}=1$ and 
\begin{align}
\label{eqn:zpmixed}
Z_{\mathcal{P}}=\frac{\frac{M_R}{Z_MZ_\mathcal{P}}+\frac{m_R}{Z_mZ_\mathcal{P}}}{M_R+m_R}\, .
\end{align}
Note that in the degenerate mass limit, we recover $Z_mZ_\gP=1$.\\

In the second case, where we take the trace with $\gamma^5$, we make use of Eq.~\ref{eq:mixedmSMOM5}, giving
\begin{align}
\left(\frac{1}{Z_{\mathcal{A}}}-\frac{\left(\frac{M_R}{Z_MZ_\mathcal{P}}+\frac{m_R}{Z_mZ_\mathcal{P}}\right)}{M_R+m_R}\right)\Tr
  \left[ \left(q\cdot\Lambda_{\mathcal{A},R}\right) \gamma^5
  \right]=&\left(1-\frac{\left(\frac{M_R}{Z_MZ_\mathcal{P}}+\frac{m_R}{Z_mZ_\mathcal{P}}\right)}{M_R+m_R}\right)\left(\Tr
  \left[-i\zeta^{-1}S_{H,R}(p_2)^{-1}-i\zeta
  S_{l,R}(p_3)^{-1}\right]\right)\, ,
\end{align}
which has solutions $Z_{\mathcal{A}}=1$ and $Z_\mathcal{P}$ as in
Eq.~\ref{eqn:zpmixed}. One can easily check that this solution is
unique.

\subsection{Finiteness of the $\zeta$ ratio}
\label{sec:finiteness-xi-ratio}

We need to show that the ratio $\zeta$ is finite since it appears
together with the renormalized propagators on the right hand sides of
Eq.~\ref{eq:mixedmSMOM3} and Eq.~\ref{eq:mixedmSMOM4} while the left
hand sides of these equations only contain renormalized vertices and
mass. For $\zeta=\frac{\sqrt{Z_l}}{\sqrt{Z_H}}$ to be finite, the
coefficient of the divergent part $Z_H$ has to be mass independent
in order to cancel with the same term in $Z_l$. We will argue that
this has to be the case order by order in perturbation theory.

The fermion propagator can be written as:
\begin{equation}
  S(p)=\frac{i}{\slashed{p}-m+i\epsilon-\Sigma(p)}\, ,
\end{equation}
where the self-energy $\Sigma(p)$ is decomposed into
\begin{equation}
  \Sigma(p)=\slashed{p}\Sigma_V(p^2)+m\Sigma_S(p^2)\, .
\end{equation}
Assuming that the theory is regulated using dimensional
regularization, let us examine all possible coefficients multiplying
the divergent terms that can appear in the self-energy at any given
order in perturbation theory. Note that $\Sigma_V(p^2)$ and
$\Sigma_S(p^2)$ are dimensionless scalars, which means the terms
appearing in the coefficient of the divergent part can only be a
function of $\ln\left(\frac{p^2}{m^2}\right)$, $\frac{p^2}{m^2}$,
$\frac{m^2}{p^2}$ or a number.

As argued in Ref.~\cite{Caswell:1981ek}, all UV divergences can be
subtracted using {\it local} counter-terms only. In other
words, the field renormalization used to remove the divergences cannot
contain terms which are functions of $\ln\left(\frac{p^2}{m^2}\right)$
and $\frac{m^2}{p^2}$, since these are non-local. The term
$\frac{p^2}{m^2}$ cannot occur either since it is IR divergent in the
limit $m\to0$ whereas we had used off-shell conditions from the
beginning and therefore do not expect any IR divergences. The only
remaining option is a coefficient proportional to 1 which has be the
same number in both the massive and massless cases since in the
absence of IR divergences $Z_H$ to reduces to $Z_l$.

Another way to argue that the divergent part of the massive
self-energy has to be mass independent is the fact that a massless
renormalization scheme removes all the divergences. Therefore $Z_H$
and $Z_l$ must have the same coefficient for their divergent terms as argued in Ref.~\cite{Weinberg:1951ss}.

\section{Lattice regularization}
\label{sec:latt-regul}

The case where chiral symmetry is broken by the regulator has been
discussed in detail in Ref.~\cite{Testa:1998ez}. Here we simply
summarise the main results, and apply them to our problem. 

When the theory is regulated on a lattice, chiral symmetry can be
broken by the regulator. 
In the case of Wilson fermions the breaking is due to the presence of higher-dimensional operators in the action, while for chiral fermions these contributions are exponentially suppressed. The net result is that symmetry breaking terms appear
in the bare Ward identities, which in turn invalidates the proof that
Noether currents do not renormalize. Assuming that the lattice
discretization reproduces the usual continuum Dirac operator in the
classical continuum limit, the variation of the action under chiral
rotations is given by higher-dimensional operators. Using the notation
introduced in Ref.~\cite{Testa:1998ez} we denote the operators
generated from the explicit symmetry breaking due to the regulator by
$X^a(x)=a O_5(x)$, where the suffix indicates that these operators are
at least of dimension 5:
\begin{equation}
  \label{eq:deltaSreg}
  - \frac{\delta S}{\delta \alpha_A(x)} = \nabla_\mu^* A^a_\mu(x)  - 
  \bar\psi(x) \left\{\tau^a, \mathcal{M}\right\} \psi(x) + X^a(x)\, ;
\end{equation}
the corresponding lattice Ward identity looks like:
\begin{align}
  \label{eq:LatWardId}
  \nabla^*_\mu \langle A^a_\mu(x) \psi(y) \bar\psi(z) \rangle = 
  2 m & \langle P^a(x) \psi(y) \bar\psi(z) \rangle + \mathrm{contact\
  terms} \nonumber \\
  + \langle X^a(x) \psi(y) \bar\psi(z) \rangle\, .
\end{align}
The current $A^a_\mu$ appearing in the Ward identity is the Noether
current associated to the symmetry transformation. In order to discuss
the symmetries of the theory in the continuum limit, the operators
appearing in Eq.~\ref{eq:LatWardId} need to be renormalized. In
particular the mixing with lower-dimensional operators, leading to
power-divergences, needs to be subtracted:
\begin{align}
  \label{eq:O5Renorm}
  O^a_{5R}(x) = Z_5 \left[
  O^a_5(x) + \frac{\overline{m}}{a} P^a(x) + \frac{Z_\gA-1}{a} \nabla^*_\mu A^a_\mu(x)
  \right]\, .
\end{align}

Ref.~\cite{Testa:1998ez} shows that these power divergences do not
contribute to the anomalous dimensions at all orders in perturbation
theory, \ie they do not depend on the renormalization scale
$\mu$. Beyond perturbation theory this result is guaranteed by the
universality of the continuum limit and the validity of the continuum
Ward identities at all scales.

In the case of chiral symmetry, the net result of the symmetry
breaking induced by the regulator is the appearance of a nontrivial
renormalization constant for the axial current:
\begin{equation}
  \label{eq:AxRen}
  A^a_{R,\mu} = Z_\gA\left(g, am\right) A^a_\mu\, ,
\end{equation}
and the renormalized current satisfies the Ward identities up to terms
that vanish when the lattice spacing goes to zero. Note that the mass
dependence in $Z_\gA$ can only enter via the dimensionless ratio
$am$. 

The same result holds if the lattice regularization preserves chiral
symmetry, but the axial current is {\em not} the Noether current
associated to the lattice symmetry.  The local currents of lattice
chiral fermions are typical examples in this category. We expect the
local currents to differ from the conserved one by irrelevant
operators. The latter need to be renormalized in order to study the
continuum limit of the Ward identities. The renormalization of the
higher-dimensional operators describing the difference between the
conserved and the non-conserved current is performed along the lines
of Eq.~\ref{eq:O5Renorm}, and yields a scale independent
renormalization constant $Z_\gA$.

\section{Numerical implementation}
\label{sec:numer-impl}

In lattice studies involving charmed and B mesons, the renormalization
of the axial current is of particular importance since it is required
to normalize correctly the matrix element entering the computation of
the decay constant. For example, the decay constants of D mesons $f_D$
and $f_{D_s}$ are determined using
\begin{align*}
\langle0|A_{cq}^\mu|D_q(p)\rangle = f_{D_q}p^\mu_{D_q},
\end{align*}
where $q=d,s$ and the axial current $A^\mu_{cq} = \bar{c}\gamma_\mu
\gamma_5 q$ has to be renormalized. Since the quark content contains a
heavy and a light quark, we can use the mass-non-degenerate mSMOM
scheme introduced in Sec.~\ref{sec:mass-non-degenerate-scheme}. The
renormalization conditions in Euclidean space are specified in
App.~\ref{sec:MinktoEuc}. Our aim is to extract the axial current renormalization $Z_\mathcal{A}$ for the mixed heavy-light vertex function. We start by writing all the ingredients needed before giving the final answer. The field renormalizations $Z_l$ and $Z_H$ are computed
using SMOM and mSMOM schemes respectively. If the {\it local} axial current is simulated on the lattice, the corresponding renormalization factor, $Z_{\text{A}^\text{local}}$, for the heavy-heavy and light-light vertex functions can be extracted by taking appropriate ratios of the respective local and conserved hadronic expectations values. Note that the correlations functions of the local and
conserved axial currents only differ by finite contributions which
vanish in continuum limit. 

Here we will now take the assumption that both quarks are constructed with chiral fermion actions, for which an explicit representation of their partially conserved, point split, axial current is available \cite{Blum:2000kn,Blum:2014tka}. We will use this to renormalize the mass degenerate local axial current bilinear operators via the WI as a component in our numerical strategy to determine the renormalization of the mixed axial current. For domain wall fermions $Z_\gA^\text{local}$ is obtained by fitting the following to a constant \cite{Blum:2000kn,Blum:2014tka},
\begin{align}
  \label{eqn:za_local}
  Z_\gA^\text{local}= \frac{1}{2} \left[ \frac{C(t-1/2)+C(t+1/2)}{2L(t)} + \frac{2 C(t+1/2)}{L(t-1)+L(t+1)} \right], 
\end{align}
where 
\begin{align}
  \label{eq:pt-split}
&  C(t+1/2) = \sum_{\bf x}\langle A_0^\text{cons}({\bf x},t)P({\bf 0},0)\rangle \ , \\
& \ \ \ \ \ \ \ \ L(t) = \sum_{\bf x}\langle A_0^\text{local}({\bf x},t)P({\bf 0},0)\rangle.
\end{align}
with $P$ being a pseudoscalar state. To obtain $Z_M$, we use  the mSMOM renormalization condition Eq.~\ref{eq:mSMOM2-Euc} to write
\begin{align}
  \label{eq:zmheavyheavy}
  Z_M=\frac{Z_H^{-1}}{12M}\left\{\left. \Tr \left[
  S(p)^{-1}\right] \right|_{p^2=-\mu^2}
  + \frac12 Z_\text{A}\left. \Tr \left[\left(iq\cdot \Lambda_{\gA}\right) \gamma_5
  \right]\right|_\sym \right\}\, .
\end{align}
where $Z_\gA$ is the renormalization constant for the heavy-heavy local current, if that is chosen, and is computed as in Eq.~\ref{eqn:za_local}. The trace of the bare vertex functions and the propagators with an
appropriate projector is numerically evaluated on the lattice. Similarly for $Z_m$, which is obtained from the SMOM scheme and the corresponding value of $Z_A$ for the light-light current. The renormalization constant for
the mass degenerate pseudoscalar density, $Z_\gP$ which can be obtained
using Eq.~\ref{eq:mSMOM1-Euc} and Eq.~\ref{eq:mSMOM5-Euc} in the mSMOM
scheme:
\begin{align}
  \label{eq:zpheavyheavy}
  Z_\gP=\frac{i}{p^2}\frac{ \left. \Tr \left[i S(p)^{-1}
  \pslash\right] \right|_{p^2=\mu^2}}
  {\Tr \left. \left[ 
  \Lambda_{\gP} \gamma_5 \right] \right|_\sym} \, .
\end{align}

Now, we can write down the equation which allows us to extract $Z_\mathcal{A}$. Recall that curly letters refer to heavy-light mixed vertices. From the renormalization conditions stated in Eq.~\ref{eq:mSMOM4-Euc} and Eq.~\ref{eq:mixedmSMOM4} we have

\begin{align}
  \label{eq:numaxialratio}
\left(  \frac{C_{\mathcal{A}(Mm)}+C_{Mm\mathcal{P}} }
          {\Delta_{H-L}}  \right)_{\text{mixed}}  =1=
     \Big(C_{\gA(MM)}+C_{M\gP} \Big) C_{\gA(mm)}    \ ,
\end{align}

where the numerator of the left hand side contains the heavy-light mixed vertex functions
\begin{align}
C_{\mathcal{A}(Mm)} = & \lim_{\substack{m_R\to0 \\ M_R\to\bar{m}}} \frac{1}{12 q^2} \Tr \left. \left[q
          \cdot \Lambda_{\mathcal{A},R}\gamma_5 \qslash\right]\right|_\sym\, \ , 
 \end{align}
\begin{align}
C_{Mm\mathcal{P}} =  \lim_{\substack{m_R\to0 \\ M_R\to\bar{m}}} \frac{1}{12 q^2} \Tr \left. \left[(M_R+m_R)
      \Lambda_{\mathcal{P},R} \gamma_5 \qslash\right]\right|_\sym\, ,
  \end{align}
 and the difference between the inverse propagators 
  \begin{align}
\Delta_{H-L}= \lim_{\substack{m_R\to0 \\ M_R\to\bar{m}}} \frac{1}{12 q^2}\Tr  \left[ \left(
          +i\gamma^5\zeta^{-1}S_{H,R}(p_2)^{-1}+i\zeta S_{l,R}(p_3)^{-1}\gamma^5 \right)
          \gamma_5\ \qslash \right]=\frac{1}{2}\left(\zeta^{-1}+\zeta\right).
  \end{align}

On the right hand side of Eq.~\ref{eq:numaxialratio} we have the heavy-heavy vertex functions,
\begin{align}
  C_{\gA(MM)} = \lim_{M_R\to\bar{m}} \frac{1}{12 q^2} \Tr \left. \left[q
    \cdot \Lambda_{\gA,R}\gamma_5 \qslash\right]\right|_\sym\, \ ,
\end{align}

\begin{equation}
  C_{M\gP} = \mlim \frac{1}{12 q^2} \Tr \left. \left[2M_R
    \Lambda_{\gP,R} \gamma_5 \qslash\right]\right|_\sym\, ,
\end{equation}
and the light-light vertex function
\begin{align}
  C_{\gA(mm)} = \lim_{m_R\to0} \frac{1}{12 q^2} \Tr \left. \left[q
    \cdot \Lambda_{\gA,R}\gamma_5 \qslash\right]\right|_\sym\, \ .
\end{align}
The quantity $\zeta$ appearing in $\Delta_{H-L}$ is computed using the renormalization conditions for the light and heavy fields Eq.~\ref{eq:mSMOM1-Euc} and taking the ratio: 
\begin{align}
  \label{eq:numericalxi}
  \zeta=\left(\frac{\left. \Tr \left[i S_l(p)^{-1}
  \pslash\right] \right|_{p^2=\mu^2}}{\left. \Tr \left[i S_H(p)^{-1}
  \pslash\right] \right|_{p^2=\mu^2}} \right)^{1/2}\, .
\end{align}
We rewrite the renormalized quantities in terms of the bare ones. Note that the aim is to extract $Z_\mathcal{A}$. On the left hand side of Eq.~\ref{eq:numaxialratio} we have
\begin{align}
  \label{eq:numeratornumericalxi}
  Z_H^{-1/2}Z_l^{-1/2}\left(\Tr \left. \left[ \left(Z_\mathcal{A}\ q
  \cdot \Lambda_{\mathcal{A}} +(Z_MM+Z_mm)Z_\mathcal{P}\Lambda_{\mathcal{P}} \right)
  \gamma_5 \ \qslash \right] \right|_\sym\right),
\end{align}
with $Z_l$ and $Z_H$ are already computed using SMOM and mSMOM schemes respectively, together with $\Delta_{H-L}$ which we have computed using Eq.~\ref{eq:numericalxi}.

Let us now focus on the right hand side of Eq.~\ref{eq:numaxialratio},
\begin{align}
  Z_H^{-1}Z_l^{-1}\left. \Tr \left. \left[ \left(Z_\gA \ q
  \cdot \Lambda_{\gA} + Z_MZ_\gP \ 2M \Lambda_{\gP} \right)
  \gamma_5 \ \qslash \right] \right|_\sym \right|_{\text{HH}}
  \left. \Tr \left. \left[ \left(Z_\gA \ q
  \cdot \Lambda_{\gA,R}  \right)
  \gamma_5 \ \qslash \right] \right|_\sym \right|_{\text{ll}}\, .
\end{align}
Therefore, all the quantities appearing in Eq.~\ref{eq:numaxialratio} are known apart from two, $Z_\mathcal{A}$ which is the main quantity we are looking for and $Z_\mathcal{P}$, which are yet to be extracted. They can both be obtained by solving the set of simultaneous equation using Eq.~\ref{eq:numaxialratio} and the renormalization condition for the pseudoscalar Eq.~\ref{eq:mixedmSMOM5}:
\begin{equation}\label{eq:casesolvezpza}
\begin{cases}
C_\mathcal{A} Z_\mathcal{A}+C_\mathcal{P} Z_\mathcal{P}=C \ ,\\
C^\prime_\mathcal{A} Z_\mathcal{A}+C^\prime_\mathcal{P} Z_\mathcal{P}=C^\prime \ ,
\end{cases}
\end{equation}
with
\begin{equation}
C_\mathcal{A}=  Z_H^{-1/2}Z_l^{-1/2}\left(\Tr \left. \left[ \left(q
  \cdot \Lambda_{\mathcal{A}}  \right)
  \gamma_5 \ \qslash \right] \right|_\sym\right)\frac{2}{\zeta^{-1}+\zeta} \ ,
\end{equation}

\begin{equation}
C_\mathcal{P} =  Z_H^{-1/2}Z_l^{-1/2}\left(\Tr \left. \left[ \left((Z_MM+Z_mm)Z_\mathcal{P}\Lambda_{\mathcal{P}} \right)
  \gamma_5 \ \qslash \right] \right|_\sym\right)\frac{2}{\zeta^{-1}+\zeta} \ ,
\end{equation}

\begin{equation}
C = \Big(C_{\gA(MM)}+C_{M\gP} \Big) C_{\gA(mm)} \ .
\end{equation}
where all the ingredients in $C$ have already been computed. Together with,
\begin{equation}
C^\prime_\mathcal{A}= - \left. \Tr \left[\left(iq\cdot \Lambda_{\mathcal{A}}\right) \gamma_5
                                               \right]\right|_\sym \ ,
\end{equation}

\begin{equation}
C^\prime_\mathcal{P}= \frac{1}{12 i} \Tr \left. \left[ 
                                                 \Lambda_{\mathcal{P}} \gamma_5 \right] \right|_\sym \ ,
\end{equation}

\begin{equation}
C^\prime =   \frac{1}{12(M_R+m_R)}\left\{\left. \Tr \left[
  S_{H}(p)^{-1}\right] \right|_{p^2=-\mu^2} + \left. \Tr \left[
  S_{l}(p)^{-1}\right] \right|_{p^2=-\mu^2} \right\} \ .
\end{equation}
Putting then all together, Eq.~\ref{eq:casesolvezpza} is solved to obtain $Z_\mathcal{P}$ and $Z_\mathcal{A}$.\\                 

The exploration of the
details of the numerical implementation is deferred to forthcoming work.

\section{Conclusions}
\label{sec:conclusions}

We have developed a mass dependent renormalization scheme, RI/mSMOM,
for fermion bilinear operators in QCD with non-exceptional momentum
kinematics similar to the standard RI/SMOM scheme. In contrast to
RI/SMOM where the renormalization conditions are imposed at the chiral
limit, this scheme allows for the renormalization conditions to be set
at some mass scale $\overline{m}$, which we are free to choose. In the
limit where $\overline{m}\to0$, our scheme reduces to SMOM.  Using a
mass dependent scheme for a theory containing massive quarks has the
benefit of preserving the continuum WI by taking into account terms of
order $m/\mu$, which would otherwise violate the WI when a massless
scheme is used. We have shown that the WIs for the case of both
degenerate and non-degenerate masses are satisfied non-perturbatively,
giving $Z_V=1$ and $Z_A=1$. In order to gain a better understanding of
the properties of the mSMOM scheme we have performed an explicit
one-loop computation in perturbation theory using dimensional
regularisation.

\acknowledgements We are indebted to Claude Duhr for his help with the
technical aspects of massive one-loop computations and the use of his
Mathematica package {\ttfamily PolyLogTools}.  AK is thankful to
Andries Waelkens and Einan Gardi for helpful discussions regarding the
perturbative calculations. LDD is supported by STFC, grant
ST/L000458/1, and the Royal Society, Wolfson Research Merit Award,
grant WM140078. AK is supported by SUPA Prize Studentship and
Edinburgh Global Research Scholarship.  LDD and AK acknowledge the
warm hospitality of the TH department at CERN, where part of this work
has been carried out. We are grateful to Andreas J\"uttner, Martin L\"uscher, Guido Martinelli, Agostino Patella and Chris Sachrajda for
comments on early versions of the manuscript.

\appendix

\section{Conventions}
\label{sec:conventions}

Let us summarise here the conventions used in this work.

\begin{itemize}
\item  The fermion propagator in position space is 
\begin{equation}
S(x_3-x_2)=\langle\psi(x_3)\bar{\psi}(x_2)\rangle,
\end{equation}
and the Fourier convention we use is
\begin{equation}\label{eq:FTmink}
S(p)=\int d^4x e^{ip.x}S(x).
\end{equation}
 The fermion propagator in momentum space is written as
\begin{equation}
S(p)=\frac{i}{\slashed{p}-m+i\epsilon-\Sigma(p)}\, ,
\end{equation}
and the fermion self-energy $\Sigma(p)$ is decomposed into
\begin{equation}
\Sigma(p)=\slashed{p}\Sigma_V(p^2)+m\Sigma_S(p^2)\, .
\end{equation}

\item The gluon propagator in Feynman gauge is 
\begin{equation}
\frac{-ig^{\mu\nu}}{k^2+i\epsilon}\, .
\end{equation}

\item Note that the one-loop self-energy $\Sigma(p)$ in this convention is 
\begin{equation}
-i\Sigma(p)=-i g^2 C_2(F)\int\frac{\gamma_\alpha(\slashed{p}-k+m)\gamma^\alpha}{k^2\left[(p-k)^2-m^2\right]}\, .
\end{equation}

\item The basis of the Clifford algebra is chosen to be $\Gamma=1 (S) , i\gamma^5 (P), \gamma^\sigma (V),
  \gamma^\sigma\gamma^5 (A),
  \sigma^{\mu\nu}=\frac{i}{2}\left[\gamma^\mu,\gamma^\nu\right] (T)$.

\item The vertex function in position space is 
\begin{equation}
G^a_O(x_3-x,x_2-x)=\langle \psi(x_3) O^a_\Gamma(x)\bar{\psi}(x_2)\rangle
\end{equation}
where we have used translational invariance and $O^a_\Gamma=\bar\psi \Gamma\tau^a\psi$ is a flavor non-singlet
fermion bilinear operator.

\end{itemize}

\section{Methods for massive one-loop computations}
\label{sec:methods-massive-one}

The 1-loop diagram in the perturbative calculation of the vertices corresponds to the following integral:
\begin{align}
 \Lambda_\Gamma^{(1)} = -ig^2C_2(F)\int_k \frac{\gamma_\mu[\slashed{p}_3-\slashed{k}+m]\Gamma[\slashed{p}_2-\slashed{k}+m]\gamma^\mu}{k^2[(p_2-k)^2-m^2][(p_3-k)^2-m^2]},
\end{align}
where $\Gamma=\gS,\gP,\gV,\gA$. 

\begin{figure}[H]
  \centering
\includegraphics[width=0.2\textwidth]{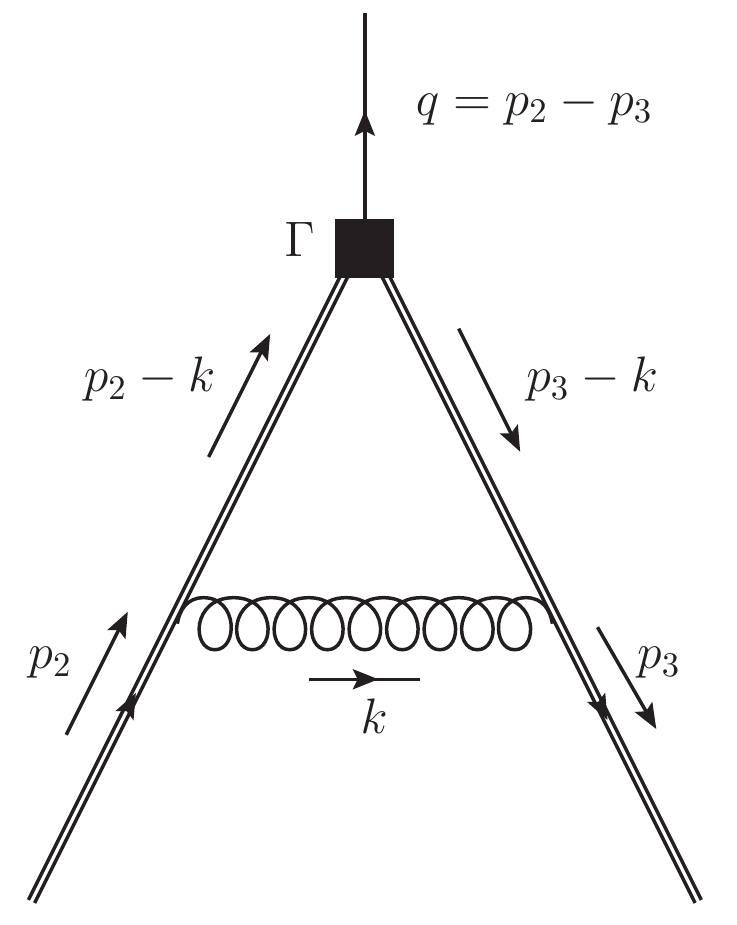}
  \caption{Diagram representing the non-amputated vertex function at 1-loop in perturbative QCD.}
  \label{fig:kin}
\end{figure}

The scalar, vector and tensor parts of the above integral are then extracted and all written in terms of scalar integrals. Then, one needs to compute the master integrals and use them to calculate each vertex $ \Lambda_\Gamma^{(1)}$. The loop integration is a standard computation, while for the integration over the Feynman parameters we have used certain techniques which have been developed in the past few years, see Ref.~\cite{Smirnov:2006ry, Smirnov:2008iw, Chavez:2012kn}.

\subsection{The scalar triangle integral}
\label{sec:scal-triangle-integr}

It is worthwhile to discuss one integral in detail, in order to
illustrate the techniques that are used in massive calculations; all
computations of massive diagrams in this work follow the same
logic. The typical scalar triangle is
\begin{equation}
  \label{eq:I111}
  I_{111} = g^2 \int_k \frac{1}{k^2} \frac{1}{(p_2-k)^2-m^2} \frac{1}{(p_3-k)^2-m^2}\, .
\end{equation}
Introducing as usual a set of Feynman parameters $x_1,x_2,x_3$, the
integral can be recast in the following form:
\begin{equation}
  \label{eq:I111Feyn}
  I_{111} = g^2 \Gamma(3) \int_k \int_0^1 \left(\displaystyle
    \prod_{i=1}^3 \mathrm{d}x_i\right) \delta\left(1-\sxi\right)
  \frac{1}{\left(x_1 k^2 + x_2 \left[(p_2-k)^2-m^2\right] + x_3
    \left[(p_3-k)^2-m^2\right] \right)^3}\, .
\end{equation}
Performing standard manipulations with Feynman parameters, and
performing a Wick rotation to Euclidean space yields:
\begin{equation}
  \label{eq:I111simpl}
  I_{111} = -i g^2 \Gamma(3) \int_0^1 \left(\displaystyle
    \prod_{i=1}^3 \mathrm{d}x_i\right) \delta\left(1-\sxi\right) 
  \frac{1}{(x_1+x_2+x_3)^3} \int_\ell
  \frac{1}{\left(\ell^2+M^2\right)^3}\, ,
\end{equation}
where we introduced the function 
\begin{equation}
  \label{eq:M2def}
  M^2 = \left(\frac{x_2 p_2 + x_3 p_3}{x_1+x_2+x_3}\right)^2 + 
  \frac{x_2+x_3}{x_1+x_2+x_3} \left(\mu^2+m^2\right)\, ,
\end{equation}
which is obtained by evaluating the square of the four-momenta at the
symmetric renormalization point. 

The loop integral can now be performed in closed form in $D$
dimensions; in this particular case the integral is finite, and the
limit $\epsilon \to 0$ is not singular. Singularities appear as poles
in $1/\epsilon$, and are treated as in the massless case. Here we want
to focus on the integral over the Feynman parameters. After the
loop integral is performed, the integral reduces to 
\begin{equation}
  \label{eq:FeynComp}
  I_{111} = -i \frac{\alpha}{4\pi} \int_0^1 \left(\displaystyle
    \prod_{i=1}^3 \mathrm{d}x_i\right) \delta\left(1-\sxi\right) 
  \frac{1}{(x_1+x_2+x_3)^3}  \frac{1}{M^2}\, .
\end{equation}
The denominator in the integrand can be expressed as
\begin{equation}
  \label{eq:FeynParDen}
  \mu^2 (x_1+x_2+x_3) \left[\
    x_2 x_3 + x_1 x_2 + x_1 x_3 + u \left(
      x_1 x_2 + x_1 x_3 + x_2^2 + x_3^2 + 2 x_2 x_3
      \right) 
    \right]\, ,
\end{equation}
where we have introduced $u=m^2/\mu^2$.
Using the Cheng-Wu theorem Ref.~\cite{Smirnov:2006ry}, applied to the case where we choose the
constraint to be $\delta(1-x_3)$, two integrations over the Feynman
parameters can be easily done, yielding
\begin{equation}
  \label{eq:OnlyOneFeyn}
  I_{111} = -i \frac{\alpha}{4\pi} \frac{1}{\mu^2} \int_0^\infty
  \mathrm{d}x_2\  
  \frac{-\log\left[-u (x_2+1)^2 - x_2\right] + 
    \log\left[-(x_2+1) (u+1) \right] +
    \log(x_2+1)}{x_2(x_2+1)+1} \, .
\end{equation}
Note that this integral can be readily computed numerically for the
case where $m=0$. The result of the numerical integration of the above integral is 2.34239 which agrees with the number
quoted in Ref.~\cite{Sturm:2009kb}.

For our purposes the analytic expression for $I_{111}$ as a function
of the mass is actually desirable. With a change of integration
variable
\[
x \mapsto y\, ,\quad x=\frac{y}{1-y}
\]
the problem is reduced to an integral that can be computed explicitly: 
\begin{equation}
  \label{eq:FinalInt}
  I_{111} = i \frac{\alpha}{4\pi} \frac{1}{\mu^2} \int_0^1
  \mathrm{d}y \frac{\log(\frac{y}{1-y}-n_1) + \log(n_1\frac{y}{1-y}-1)
    - \log(n_1) - 2 \log(\frac{y}{1-y}+1) + \log(u) - \log(u+1)}{(y +
    (y-1) d_1) (y + \frac{y-1}{d_1})}\, ,
\end{equation}
where $d_1=\frac12 \left( -1 + i \sqrt{3} \right), n_1=\frac12 \left(
  -2 -1/u - \sqrt{1/u^2 + 4/u} \right)$. The final result is a lengthy
expression, which we report for completeness,
\begin{align}
  \label{eq:IFinal}
  I_{111} = \frac{\alpha}{4\pi} \frac{1}{\mu^2} \frac{1}{\sqrt{3}} 
  \Bigg\{& i \frac{\pi}{3} (-2i\pi - 2\log(1 + u)) \nonumber \\
         &+ \log\left[-\frac{2u+1 -\sqrt{1+4u}}{2}\right] 
           \log \left[ \frac{4 + (i \sqrt{3} -1) (1-\sqrt{4u+1}) }{4 -
           (i \sqrt{3} + 1) (1-\sqrt{4u+1}) } \right] \nonumber \\
         &+ \log\left[-\frac{(1+\sqrt{4u+1})^2}{4}\right]
           \log \left[ \frac{4 + (i \sqrt{3} -1) (1+\sqrt{4u+1}) }{4 -
           (i \sqrt{3} + 1) (1+\sqrt{4u+1}) } \right] \nonumber \\
         &+ 2 \li\left[\frac{4u}{4u-\left(i \sqrt{3} - 1\right)
           \left(1+\sqrt{4u+1}\right)}\right]\
         - 
         \li \left[\frac{4u}{4u+\left(i \sqrt{3}+1\right)
         \left(1+\sqrt{4u+1}\right)}\right] \nonumber \\ 
         &+ \li
           \left[\frac{4u+2+2\sqrt{4u+1}}{4u+\left(i\sqrt{3}+1\right)
           \left(1+\sqrt{4u+1}\right)}\right] -
           \li \left[\frac{4u+\left(i\sqrt{3} + 1\right)\left(1
           +\sqrt{4u+1} \right)}{4(1+u)}\right] 
           \Bigg\}\, . 
\end{align}
As a partial check of our massive computation, the limit $u \to 0$ of the
expression above is numerically evaluated, and shown to reproduce again the value
2.34391 from Ref.~\cite{Sturm:2009kb}. In the paper we denote 
\begin{equation}
  \label{eq:numericC0}
I_{111}=-\frac{i\alpha}{4\pi}\frac{1}{\mu^2}C_0\left(\frac{m^2}{\mu^2}\right) \, ,
\end{equation}
so that $C_0|_{m=0}=2.34391$.

\section{Minkowski to Euclidean convention}
\label{sec:MinktoEuc}
The renormalization conditions stated in the paper are set in Minkowski space. Here, we state our conventions for going from Minkowski to Euclidean space and use these to construct the ratio in Sec.~\ref{sec:numer-impl} for numerical implementation. We take
\begin{align}
x^{0\text{M}}=-ix_4^\text{E} \ \ , \ \ x^{i\text{M}}=x_i^\text{E} \ ,
\end{align}
which means $x_i=-x_i^{\text{E}}$ and we do not distinguish between upper and lower indices in Euclidean space. 

Similarly for momentum $k^\mu$ we have
\begin{align}
k^{0\text{M}}=-ik_4^\text{E} \ \ , \ \ k^{i\text{M}}=k_i^\text{E} \ .
\end{align}
The relation for the vector potential becomes
\begin{align}
A^{0\text{M}}=iA_4^\text{E} \ \ , \ \ A^{i\text{M}}=-A_i^{E} \ .
\end{align}
Therefore the covariant derivative in Minkowski space
\begin{align}
D_\mu=\partial_\mu+ig\mathscr{A}_\mu \ ,
\end{align}
maps to 
\begin{align}
D^{0\text{M}}=iD_4^\text{E}\ \ , \ \ D^{i\text{M}}=-D_i^\text{E} \ , 
\end{align}
and the Euclidean covariant derivative becomes
\begin{align}
D^\text{E}_\mu=\partial^\text{E}_\mu+ig\mathscr{A}_\mu^\text{E} \ .
\end{align}
The gamma matrices map in the following way:
\begin{align}
\gamma^{0\text{M}}=\gamma_4^\text{E} \ \ , \ \ \gamma^{1,2,3\text{ M}}=i\gamma_{1,2,3}^\text{E} \ .
\end{align}
For convenience we also take 
\begin{align}
\psi^\text{M}=\psi^\text{E} \ \ , \ \ \overline{\psi}^{\text{M}}=\overline{\psi}^{E} \ .
\end{align}
The fermionic part of the action in Euclidean space becomes
\begin{align}
S^\text{E}[\overline{\psi},\psi]=\int d^4x^\text{E} \ \overline{\psi}^\text{E}\bigg[\gamma_\mu^{\text{E}} D_\mu^{E}+m\bigg]\psi^\text{E} ,
\end{align}
The renormalization condition in Euclidean space are:
\begin{align}
  \label{eq:mSMOM1-Euc}
  \mlim &\left. \frac{1}{12 p^2_\text{E}} \Tr \left[iS_R^\text{E}(p)^{-1}
     \pslash^\text{E}\right] \right|_{p^2_\text{E}=\mu^2}=-1\, , \\
  \label{eq:mSMOM2-Euc}
  \mlim & \frac{1}{12 M_R} \left\{\left. \Tr \left[
     S^\text{E}_R(p)^{-1}\right] \right|_{p^2=\mu^2}
     + \frac12 \left. \Tr \left[\left(iq\cdot \Lambda^\text{E}_{\gA,R}\right) \gamma_5
     \right]\right|_\sym \right\}=1\, , \\
   \label{eq:mSMOM3-Euc}
  \mlim & \frac{1}{12 q^2} \Tr \left. \left[ \left(q
          \cdot \Lambda_{\gV,R} \right) \qslash \right] \right|_\sym =
          1\, , \\
   \label{eq:mSMOM4-Euc}
  \mlim & \frac{1}{12 q^2} \Tr \left. \left[ \left(q
          \cdot \Lambda_{\gA,R} + 2M_R \Lambda_{\gP, R} \right)
          \gamma_5 \qslash \right] \right|_\sym = 1\, ,\\
   \label{eq:mSMOM5-Euc}
  \mlim &\frac{1}{12 i} \Tr \left. \left[ 
          \Lambda_{\gP,R} \gamma_5 \right] \right|_\sym = 1\, .
\end{align}
The conditions are now defined at the symmetric point, 
\begin{align}
 p_2^2 = p_3^2 = q^2 = \mu^2.
\end{align}
The RI/mSMOM scheme for the heavy-light mixed case in Euclidean space now reads:
\begin{align}
  \label{eq:mixedmSMOM3}
  \lim_{\substack{m_R\to0 \\ M_R\to\overline{m}}} & \frac{1}{12 q^2} \Tr \left. \left[ \left(q
                                               \cdot \Lambda_{\mathcal{V},R} + (M_R-m_R)\Lambda_{\mathcal{S},R} \right) \qslash \right] \right|_\sym =
                                               \lim_{\substack{m_R\to0 \\ M_R\to\overline{m}}}  \frac{1}{12 q^2} \Tr  \left[ \left(-i\zeta^{-1}S_{H,R}(p_2)^{-1}+i\zeta S_{l,R}(p_3)^{-1}\right)
  \qslash \right] 
  \, , \\
  \label{eq:mixedmSMOM4}
  \lim_{\substack{m_R\to0 \\ M_R\to\overline{m}}}  & \frac{1}{12 q^2} \Tr \left. \left[ \left(q
                                                \cdot \Lambda_{\mathcal{A},R}+ (M_R+m_R)  \Lambda_{\mathcal{P}, R}\right)
                                                \gamma_5 \qslash \right] \right|_\sym = \nonumber \\
&  \ \ \ \ \ \ \ \ \ \ \ \ \ \ \ \ \ \ \ \ \ \ \ \ \ \ \ \ \ \ \ \ \ \ \ \ \ \ \ \ \ \ \ \ \ \ \ \ \ \ \ \ \ \lim_{\substack{m_R\to0 \\ M_R\to\overline{m}}}    \frac{1}{12 q^2} \Tr  \big[ \big(
  +i\gamma^5\zeta^{-1}S_{H,R}(p_2)^{-1} +i\zeta S_{l,R}(p_3)^{-1}\gamma^5 \big)
  \gamma_5 \qslash \big] \, ,\\
  \label{eq:mixedmSMOM5}
  \lim_{\substack{m_R\to0 \\ M_R\to\overline{m}}}   &\frac{1}{12 i} \Tr \left. \left[ 
                                                 \Lambda_{\mathcal{P},R} \gamma_5 \right] \right|_\sym 
                                                 = \lim_{\substack{m_R\to0 \\ M_R\to\overline{m}}}  \Bigg\{
  \frac{1}{12(M_R+m_R)}\left\{\left. \Tr \left[
  \zeta^{-1} S_{H,R}(p)^{-1}\right] \right|_{p^2=-\mu^2}
  + \frac12 \left. \Tr \left[\left(iq\cdot \Lambda_{\mathcal{A},R}\right) \gamma_5
  \right]\right|_\sym \right\}+  \nonumber \\
                                             & \ \ \ \ \ \ \ \ \ \ \ \ \ \ \ \ \ \ \ \ \ \ \ \ \ \ \ \ \ \ \ \ \ \ \ \ \ \ \ \  \frac{1}{12(M_R+m_R)}\left\{\left. \Tr \left[
                                               \zeta S_{l,R}(p)^{-1}\right] \right|_{p^2=-\mu^2}
                                               + \frac12 \left. \Tr \left[\left(iq\cdot \Lambda_{\mathcal{A},R}\right) \gamma_5
                                               \right]\right|_\sym \right\}
                                               \Bigg\}\, .
\end{align}

\end{document}